\documentclass{IEEEtran}
\pdfoutput=1
\ifCLASSINFOpdf
\else
   \usepackage[dvips]{graphicx}
\fi
\usepackage{url}

\usepackage{amsmath}

\DeclareFontFamily{U}{mathx}{}
\DeclareFontShape{U}{mathx}{m}{n}{<-> mathx10}{}
\DeclareSymbolFont{mathx}{U}{mathx}{m}{n}
\DeclareMathAccent{\widehat}{0}{mathx}{"70}
\DeclareMathAccent{\widecheck}{0}{mathx}{"71}

\usepackage{multirow}
\usepackage{pifont}
\usepackage{amssymb}
\usepackage{graphicx,color}
\usepackage{mathrsfs}
\usepackage{bm}
\usepackage{amsfonts,amssymb}
\usepackage{extarrows}
\usepackage{graphicx}
\usepackage{algorithm}
\usepackage{algorithmic}
\usepackage{setspace}
\usepackage{array}
\usepackage{epstopdf}
\usepackage{subfigure}
\usepackage{caption}
\usepackage{float}

\usepackage{cite}
\usepackage{url}
\usepackage{stfloats}
\DeclareMathOperator*{\argmax}{arg\,max}
\date{}
\begin{document}

\title{Environment Reconstruction based on Multi-User Selection and Multi-Modal Fusion in ISAC}
\author{Bo Lin, Chuanbin Zhao, Feifei Gao, \IEEEmembership{Fellow, IEEE}, and Geoffrey Ye Li, \IEEEmembership{Fellow, IEEE}
\thanks{
B. Lin and F. Gao are with the Department of Automation, Tsinghua University, State Key Lab of Intelligent Technologies and Systems, Tsinghua University, State Key for Information Science and Technology (TNList), Beijing 100084, P. R. China (e-mail: feifeigao@ieee.org; linb20@mails.tsinghua.edu.cn).

C. Zhao is with the State Key Laboratory of Intelligent Technologies and Systems, Beijing National Research Center for Information Science and Technology (BNRist), Department of Automation, Institute for Artificial Intelligence Tsinghua University (THUAI), Tsinghua University, Beijing 100084, China, and also with the Senior Engineer China Telecom Corporation Sichuan Branch, Chengdu 610000, China (e-mail: zcb23@mails.tsinghua.edu.cn).

Geoffrey Ye Li is with the Department of Electrical and Electronic Engineering, Imperial College London, SW7 2BU London, U.K. (e-mail:Geoffrey.Li@imperial.ac.uk).

}}
\maketitle
\thispagestyle{empty}
\begin{abstract}
Integrated sensing and communications (ISAC) has been deemed as a key technology for the sixth generation (6G) wireless communications systems. In this paper, we explore the inherent clustered nature of wireless users and design a multi-user based environment reconstruction scheme. Specifically, we first select users based on the estimation precision of channel's multipath, including the line-of-sight (LOS) and the non-line-of-sight (NLOS) paths, to enhance the accuracy of environment reconstruction. Then, we develop a fusion strategy that merges communications signalling with camera image to increase the accuracy and robustness of environment reconstruction. The simulation results demonstrate that the proposed algorithm can achieve a remarkable sensing accuracy of centimeter level, which is about 17 times better than the scheme without user selection. Meanwhile, the fusion of communications data and vision data leads to a threefold accuracy improvement over the image only method, especially under challenging weather conditions like raining and snowing.
\end{abstract}


\IEEEpeerreviewmaketitle

\section{Introduction}\label{s_introduction}

The swift progress in artificial intelligence (AI) has notably propelled the development of sensing-assisted communications technology \cite{xu2022computer,lin2024multi,chen2022computer,charan2021vision,qin2023generalized,10152012}. Various studies have employed sensing data, such as vision and radar, to improve the efficiency and quality of wireless communications.
Reversely, wireless communications systems can fulfill the dual role of sensing the surrounding environment and data transmission \cite{wei2022toward}.
The transmitted signals complexly interact with the environment during their journey to the receivers. Thus, the ultimately received signals carry a rich range of critical environment \textcolor[rgb]{0.00,0.00,0.00}{information or} data. In fact, the expansion of communication frequency and bandwidth in the coming 6G is opening new opportunities for communication-assisted sensing.

Recently, integrated sensing and communications \cite{liu2022integrated} (ISAC) has been proposed to facilitate concurrent high-speed data transmission and high-precision sensing. 
The development of ISAC gradually converges into two categories: moving target sensing \cite{luo2023integrated,chen2023simultaneous,luo2023moving,xiang2023esprit} and static environment reconstruction \cite{leitinger2019belief,wang2023bayesian,yang2023angle,kim20205g}. In \cite{chen2023simultaneous}, a simultaneous beam training and target sensing scheme \textcolor[rgb]{0.00,0.00,0.00}{has been proposed}.
In \cite{luo2023moving}, a root-MUSIC-based algorithm \textcolor[rgb]{0.00,0.00,0.00}{has been developed} to estimate the kinematic parameters of identified moving targets in a cluttered environment.
The ESPRIT-based moving target sensing method \textcolor[rgb]{0.00,0.00,0.00}{in \cite{xiang2023esprit} can achieve} super-resolution and low-complexity estimation of the targets' parameters.
On the aspect of environment reconstruction, the concept of simultaneous localization and mapping (SLAM) \textcolor[rgb]{0.00,0.00,0.00}{can} construct a comprehensive radio map based on the multipath channel state information (CSI). The belief propagation (BP) based SLAM algorithm \textcolor[rgb]{0.00,0.00,0.00}{in \cite{leitinger2019belief}} utilizes the association of specular multipath components (MPCs) with geometric features to reconstruct the environment. In~\cite{wang2023bayesian}, a Bayesian approach \textcolor[rgb]{0.00,0.00,0.00}{has been} designed for communication-driven SLAM by extracting the soft information of channel parameters. The angle-based SLAM algorithm \textcolor[rgb]{0.00,0.00,0.00}{in \cite{yang2023angle} extends} the classic BP SLAM algorithm. The multiple-model probability hypothesis density filter and map fusion routine \textcolor[rgb]{0.00,0.00,0.00}{in \cite{kim20205g} can effectively map} the radio environment.

However, \textcolor[rgb]{0.00,0.00,0.00}{existing works on reconstructing} the environment \textcolor[rgb]{0.00,0.00,0.00}{are} merely based on one single user, resulting in limited sensing information, sparse reconstruction results, and low reliability.
Actually, the advantage of communications-sensing over radar-sensing lies in the fact that the former can utilize information from the massive users in communications systems. \textcolor[rgb]{0.00,0.00,0.00}{Therefore, it has been demonstrated in \cite{sun2024integrated} that multi-user sensing can enhance the sensing performance.}
A centralized multi-user collaborative mapping and positioning approach \textcolor[rgb]{0.00,0.00,0.00}{has been proposed in \cite{chu2021vehicle}}.
The robust SLAM algorithm \textcolor[rgb]{0.00,0.00,0.00}{in \cite{yang2021enabling} extends} the classic BP-based SLAM algorithm to multi-user scenarios.
However, due to variations in user positions and user orientations, the degree of alignment between the beam scanning direction and the angle of reflection paths varies among different users, resulting in different \textcolor[rgb]{0.00,0.00,0.00}{accuracies} in environment reconstruction when only one user is used for sensing.

Nevertheless, the constructed environment point clouds are sparse due to the sparsity of the mmWave propagation paths.
Traditional method utilizes visual sensors, which can capture rich information to reconstruct the environment.
However, visual sensors have a limited detection range and \textcolor[rgb]{0.00,0.00,0.00}{therefore} are highly influenced by weather conditions.

Some studies have focused on the fusion of radar data and vision data for environment reconstruction and depth estimation. The fusion of radar and vision \textcolor[rgb]{0.00,0.00,0.00}{has been proposed in \cite{nabati2021centerfusion}} for 3D object detection. The modified encoder-decoder deep convolutional neural network (CNN) \textcolor[rgb]{0.00,0.00,0.00}{in \cite{niesen2020camera} can} fuse the camera's and radar's measurements for depth reconstruction. The geometric method \textcolor[rgb]{0.00,0.00,0.00}{in \cite{el2015radar} performs} 3D reconstruction using a panoramic microwave radar and a camera.
The point cloud reconstruction approach \textcolor[rgb]{0.00,0.00,0.00}{in \cite{yang2023mmwave} fuses} millimeter wave radar data and vision data.
In \cite{lin2020depth}, the authors explored the possibility of achieving a more accurate depth estimation by fusing monocular images and radar points using a deep neural network (DNN).

Due to the high similarity between radar systems and communications systems\cite{10251770}, it is possible to fuse the communications signals in ISAC with vision data for environment reconstruction. However, there are still several challenges when fusing ISAC and vision:
\textcolor[rgb]{0.00,0.00,0.00}{\begin{itemize}
\item[(i)] The sensing information obtained from communications \textcolor[rgb]{0.00,0.00,0.00}{is} sparser compared to radar.
\item[(ii)] Radar sensing operates in a self-transmit and self-receive manner, capturing only reflection information while communication sensing involves both direction and reflection information, which requires extra effort to identify the reflection paths.
\item[(iii)] The sensing angle range of radar sensing is well-defined while the sensing angle range of communication sensing is random.
\item[(iv)] The fusion mechanism of communications data with vision data is not yet clearly defined.
\end{itemize}}

In this paper, we reconstruct environment based on multi-user selection and multi-sensor fusion. The main contributions of this paper are as follows:
\begin{itemize}
\item We propose \textcolor[rgb]{0.00,0.00,0.00}{a} criterion to select users to enhance the accuracy of environment reconstruction.
\item We leverages the clustered nature of \textcolor[rgb]{0.00,0.00,0.00}{wireless} users to enable rich sensing information.
\item We fuse the information from ISAC and vision, and design a multi-modal fusion network (MMFN) to obtain accurate and robust environment reconstruction.
\item We adopt the meta-learning strategy to train the MMFN to guarantee the effectiveness of the MMFN across different user quantities.
\end{itemize}

The \textcolor[rgb]{0.00,0.00,0.00}{rest} of this paper is organized as follows.
Section~\ref{s_system_model} introduces the multi-user selection and multi-sensor fusion based system model.
Section~\ref{s_user_selection} presents the evaluation criterion for assessing the sensing capabilities of users and proposes the user selection algorithms.
Section~\ref{s_environment_reconstruction} designs the multi-modal fusion network for environment reconstruction.
Section~\ref{simulation} provides the dataset generation and simulation results and Section~\ref{conclusion} draws the conclusion.

\section{System Model}\label{s_system_model}
We consider an orthogonal frequency-division multiplexing (OFDM) mmWave communications system with one BS and $N$ users.
The BS is equipped with a uniform planar array (UPA) of $N_t$ antennas, and the user is equipped with a UPA of $N_r$ antennas.
Denote the set of users as $\mathcal U = \{u_{1},u_2,\cdots,u_{N}\}$.
Considering the cost of hardware deployment, both the BS and the users adopt a fully analog architecture with only one radio frequency (RF) chain.
Assume that there are $L_c$ multipath components (MPC) between the user and the BS. The wireless parameters of the $l$-th MPC provided by the wideband mmWave geometric channel model~\cite{heath2016overview} are: complex path gain $\alpha_l$, time delay $\tau_l$, azimuth angle of departure (AoD) $\phi_{t,l}$,  elevation AoD $\theta_{t,l}$, azimuth angle of arrival (AoA) $\phi_{r,l}$, and elevation AoA $\theta_{r,l}$. The channel for the $n$-th OFDM symbol is
\begin{equation}\label{channel-time}
\begin{aligned}
\mathbf H[n] = \sqrt{N_rN_t}\sum_{l=0}^{L_c-1}\alpha_l g(nT-\tau_l)\mathbf a(\phi_{r,l},\theta_{r,l})\mathbf a^*(\phi_{t,l},\theta_{t,l}),
\end{aligned}
\end{equation}
where $g(\cdot)$ is the shaping pulse, $T=\frac{1}{B}$ is the symbol period, $B$ is the bandwidth, $\mathbf a(\phi_{r,l},\theta_{r,l})$ is the array steering vector at the receiver, and $\mathbf a(\phi_{t,l},\theta_{t,l})$ is the array steering vector at the transmitter.
\textcolor[rgb]{0.00,0.00,0.00}{Denote $K$ as} the total number of OFDM subcarriers. The frequency domain channel at subcarrier $k$ is
\begin{equation}\label{channel-frequency}
\begin{aligned}
\mathbf H[k] = \sum_{n=0}^{L-1}\mathbf H[n]e^{-j\frac{2\pi k}{K}n},
\end{aligned}
\end{equation}
where $L$ is the maximum discrete-time delay of the channel.

\begin{figure}
  \centering
  \includegraphics[width=0.35\textwidth]{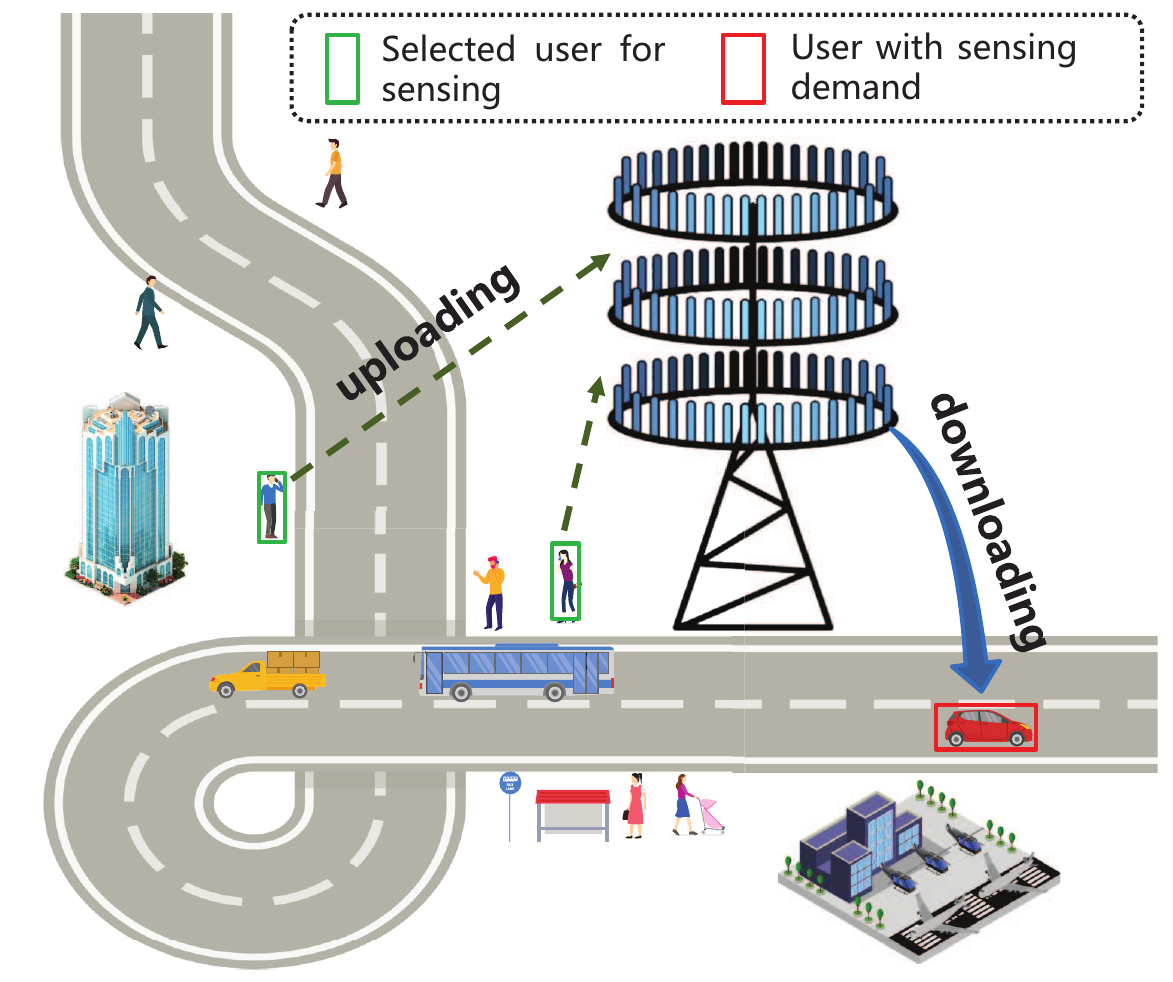}\\
  \caption{The multi-user, multi-sensor fusion system for environment reconstruction.}\label{system2}
\end{figure}

Analog beamforming is employed at both the transmitter and the receiver based on fixed beam codebooks. Denote the transmit beam codebook as ${\mathcal F_t = \{\mathbf a(\phi_{t,i}^b,\theta_{t,j}^b)|\phi_{t,i}^b \in \Phi_t^b, \theta_{t,j}^b\in\Theta_t^b\}}$ and the receive beam codebook as ${\mathcal F_r = \{\mathbf a(\phi_{r,p}^b,\theta_{r,q}^b)|\phi_{r,p}^b \in \Phi_r^b, \theta_{r,q}^b\in\Theta_r^b\}}$, where
$\Phi_t^b$ is the candidate set of transmit azimuth angle, $\Theta_t^b$ is the candidate set of transmit elevation angle, $\Phi_r^b$ is the candidate set of receive azimuth angle, and $\Theta_r^b$ is the candidate set of receive elevation angle\footnote{The angles in $\Phi_t^b$ and $\Theta_t^b$ are uniformly distributed.}. At the beam management phase\cite{parkvall2017nr} of each user, the BS and the user perform exhaustive beam sweeping among all directions in the transmit and the receive codebooks. To simplify notation, we use quad ${(i,j,p,q)}$ to represent performing beamforming with beam pair $\{\mathbf a(\phi_{t,i}^b,\theta_{t,j}^b), \mathbf a(\phi_{r,p}^b,\theta_{r,q}^b)\}$. Given a quad ${(i,j,p,q)}$, the received signal power is
\begin{equation}\label{channel-frequency}
\begin{aligned}
y_{(i,j,p,q)}=\sum_{k=0}^{K-1}|\mathbf a^H(\phi_{r,p}^b,\theta_{r,q}^b)\mathbf H[k]\mathbf a(\phi_{t,i}^b,\theta_{t,j}^b)|^2.
\end{aligned}
\end{equation}
After exhaustive beam sweeping, the user recodes the signal powers in a four dimension power map (PM) tensor {$\mathbf P_u\in~\mathbb R^{|\Phi_t^b|\cdot|\Theta_t^b|\cdot|\Phi_r^b|\cdot|\Theta_r^b|}$}, where $\mathbf P_u[i,j,p,q]=y_{(i,j,p,q)}$, and then feeds back the PM tensor to the BS.

\section{User Selection Based Environment Reconstruction}\label{s_user_selection}
\begin{figure}
  \centering
  \includegraphics[width=0.35\textwidth]{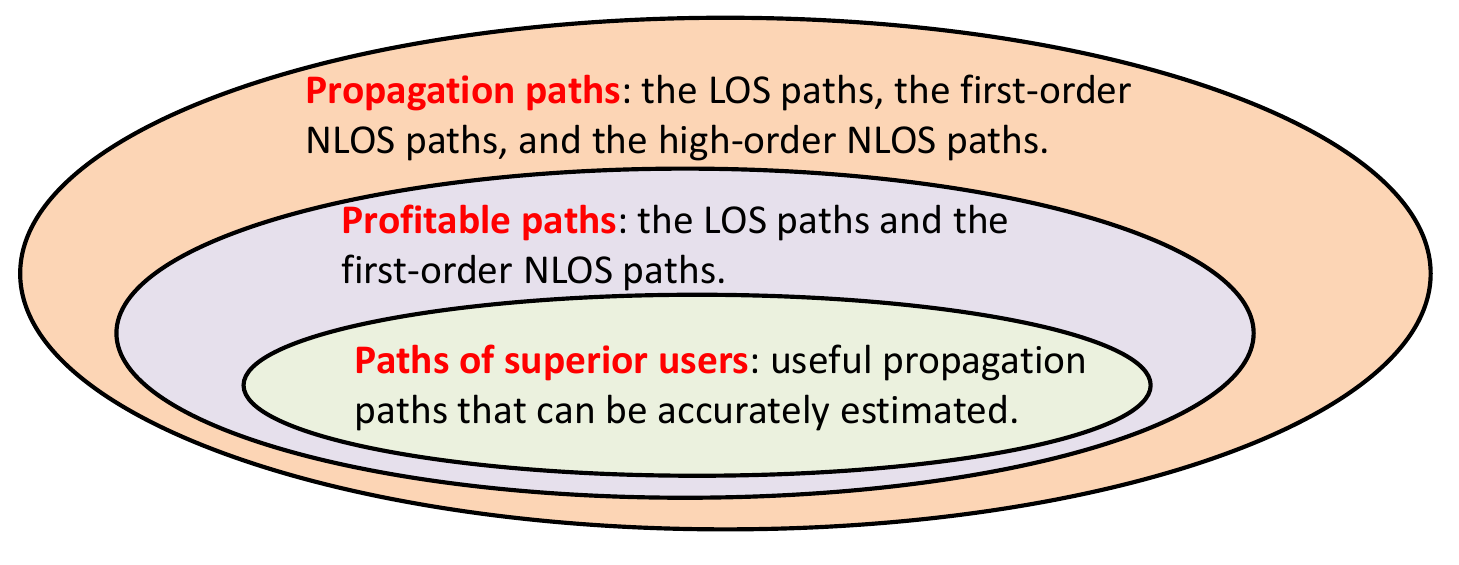}\\
  \caption{The inclusion relationship between paths.}\label{ven}
\end{figure}

The environment reconstruction is generally realized by collecting the reflection points of the first-order non-line-of-sight (NLOS) paths to a \emph{point set} $\mathcal P$\cite{mou2023millimeter}.
Specifically, we use the line-of-sight (LOS) path to calculate the location of the user, and then utilize the user's location as well as the first-order NLOS paths to calculate the reflection points.
Hence, in order to obtain accurate points, we have to precisely estimate the angles of the LOS and the first-order NLOS paths from the PM tensor, which includes the powers of the LOS path, the powers of the first-order NLOS paths, the powers of the high-order NLOS paths, and the powers of the noise.
We define \textcolor[rgb]{0.00,0.00,0.00}{the \emph{profitable paths} as} the LOS paths and the NLOS paths that are beneficial for environment reconstruction as shown in Fig. \ref{ven}.
Subsequently, we define \textcolor[rgb]{0.00,0.00,0.00}{the {superior users} as the ones} whose LOS and first-order NLOS paths can be precisely estimated from the PM tensor.
\textcolor[rgb]{0.00,0.00,0.00}{We} divide all user \textcolor[rgb]{0.00,0.00,0.00}{set} $\mathcal U$ into two sets, \textcolor[rgb]{0.00,0.00,0.00}{$\mathcal U_s$ and $\mathcal U_{n}$, according to whether they are superior users or not.}
The steps of inferring whether a user is a {superior user} are:
\textcolor[rgb]{0.00,0.00,0.00}{\begin{itemize}
\item[(i)] to extract the profitable paths that include the LOS paths and the first-order NLOS paths;
\item[(ii)] to distinguish whether the profitable paths are LOS paths or first-order NLOS paths;
\item[(iii)] to judge whether the LOS paths and the first-order NLOS paths can be accurately estimated.
\end{itemize}}

The above steps will be elaborated subsequently.

\subsection{Identify The Profitable Paths}\label{s_connection_factor}

Among the PM tensors, the powers in the directions of the propagation paths are large while the noise powers in other directions are small.
Then, we \textcolor[rgb]{0.00,0.00,0.00}{can use power as a} threshold $T_u$ to distinguish the propagation paths and the noise.
Specifically, element $\mathbf P_u[i,j,p,q]\geq T_u$ \textcolor[rgb]{0.00,0.00,0.00}{is regarding as} the propagation path while $\mathbf P_u[i,j,p,q]\textless T_u$ \textcolor[rgb]{0.00,0.00,0.00}{is regarding as} the noise.
A good threshold value $T_u$ should make the inter-class variance between the propagation path and the noise large \cite{otsu1979threshold}.
The steps to calculate the inter-class variance are as follows:
\begin{itemize}
\item[$\bullet$] \textcolor[rgb]{0.00,0.00,0.00}{Discretizing} the values in the $\mathbf P_u$ into $K$ values\footnote{$K$ is a customizable parameter. When $K$ is large, the discrete value is close to the continuous value. Then the calculated threshold will be accurate but complexity will be high.} to reduce computational complexity. 
\item[$\bullet$] \textcolor[rgb]{0.00,0.00,0.00}{Counting} the number of occurrences for each discrete power, represented as $n_k$.
\item[$\bullet$] \textcolor[rgb]{0.00,0.00,0.00}{Calculating the average power} of the propagation {paths~as}
\begin{equation}\label{m1}
m_1 = \sum_{k=1}^{T_u}n_k \cdot \sum_{k=1}^{T_u}\frac{kn_k}{\sum_{j=1}^{K}n_j}.
\end{equation}
\item[$\bullet$] \textcolor[rgb]{0.00,0.00,0.00}{Calculating the average power} of noise as
\begin{equation}\label{m1}
m_2 = \sum_{k=T_u+1}^{K}n_k \cdot \sum_{k=T_u+1}^{K}\frac{kn_k}{\sum_{j=1}^{K}n_j}.
\end{equation}
\item[$\bullet$] \textcolor[rgb]{0.00,0.00,0.00}{Calculating the average power} of fully $\bm P_u$ as
\begin{equation}\label{m2}
m_G = \sum_{k=1}^{K}\frac{kn_k}{\sum_{j=1}^{K}n_j}.
\end{equation}
\item[$\bullet$] \textcolor[rgb]{0.00,0.00,0.00}{Calculating} the inter-class variance as
\begin{equation}\label{mg}
\begin{aligned}
\sigma^2(T_u) = \frac{(m_1-m_G)^2}{\sum_{k=1}^{T_u}n_k}+\frac{(m_2-m_G)^2}{\sum_{k=T_u+1}^{K}n_k}.
\end{aligned}
\end{equation}
\end{itemize}
Then, the optimal $T_u$ \textcolor[rgb]{0.00,0.00,0.00}{should maximize} the inter-class variance as
\begin{equation}\label{threshold}
T_u^* = \argmax_{1\leq T_u\leq K}\sigma^2(T_u).
\end{equation}

\begin{figure}[t]
	\centering
	\subfigure[Original power map.]{
		\begin{minipage}[b]{0.225\textwidth}
			\includegraphics[width=0.98\textwidth]{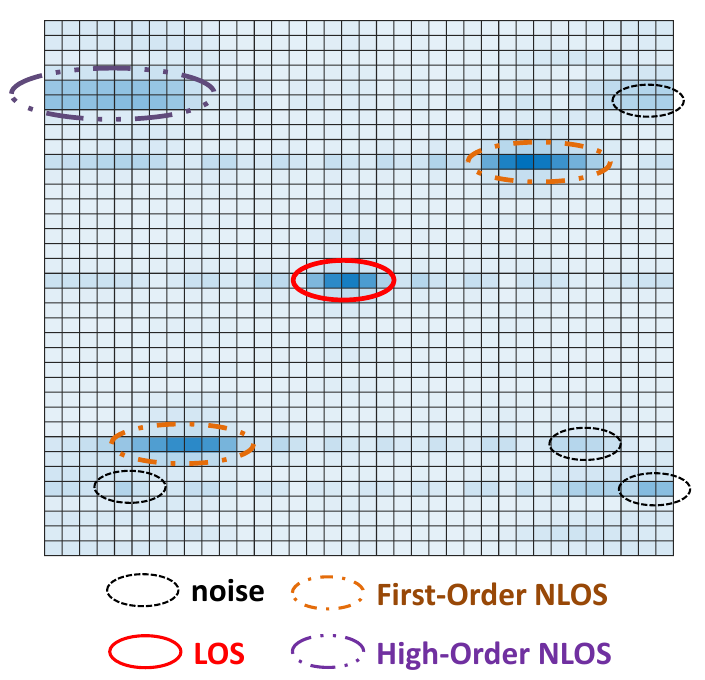}
		\end{minipage}
	}
    	\subfigure[Binarized power map.]{
    		\begin{minipage}[b]{0.225\textwidth}
   		 	\includegraphics[width=1\textwidth]{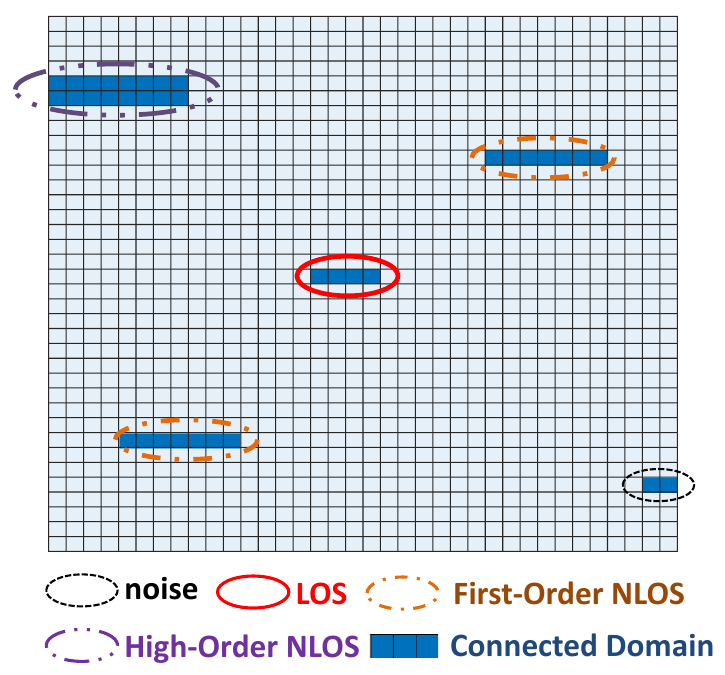}
    		\end{minipage}
		\label{fig:hor_2figs_1cap_2subcap_2}
    	}
	\caption{An example of power map.}\label{pm}
\end{figure}

Next, we mark the propagation paths by setting the elements in $\mathbf P_u$ that are greater than or equal to $T_u^*$ as ``1''; \textcolor[rgb]{0.00,0.00,0.00}{otherwise,} setting the elements in $\mathbf P_u$ as ``0'', \textcolor[rgb]{0.00,0.00,0.00}{that is,}
\begin{equation}\label{binarize}
\begin{aligned}
\mathbf P_u^b[i,j,p,q] = \textup{binarize}(\mathbf P_u[i,j,p,q],T_u^*)\\
=\left\{
\begin{aligned}
0 & , & P_u[i,j,p,q]< T_u^*, \\
1 & , & P_u[i,j,p,q]\geq T_u^*.
\end{aligned}
\right.
\end{aligned}
\end{equation}
Define the area among $\mathbf P_u^b$  whose elements are all ``1'' and are adjacent to each other as a \emph{connected domain} as shown in Fig.~\ref{pm}(b).
Denote $n_u$ connected domains of $\mathbf P_u^b$ as $\mathcal C_1^u, \mathcal C_2^u, \cdots, \mathcal C^u_{n_u}$.
Each $\mathcal{C}_k^u$ encompasses the angular coordinates of the $k$-th connected domain. The presence of a connected domain suggests the possible existence of a propagation path.
Denote the angles (AoA and AoD) of a propagation path as $(\hat i,\hat j,\hat p,\hat q)$.
Since the angles close to \textcolor[rgb]{0.00,0.00,0.00}{that} of the propagation path also yield high powers, a valid propagation path will yield high values for multiple adjacent elements in $\mathbf P_u$ as shown in Fig.~\ref{pm}(a).
Then one propagation path would yield multiple elements among the connected domain as shown in Fig.~\ref{pm}(b).
Conversely, a connected domain consisting of only one or a few elements may indicate a false alarm without a propagation path.
Moreover, among the propagation paths, the angle spreads of the high-order NLOS paths are generally larger than that of the profitable paths \cite{fleury2000first}.
Hence, as shown in Fig.~\ref{pm}(b), a connected domain consisting of relatively \textcolor[rgb]{0.00,0.00,0.00}{more} elements may indicate a \textcolor[rgb]{0.00,0.00,0.00}{higher}-order NLOS path.
Then, we define a connectivity factor to indicate the profitable path as
\begin{equation}\label{connection_factor}
\begin{aligned}
c_u^k=\textup{sign}\left(|\mathcal C_u^k|-\textup{thr}_c\right)\cdot\textup{sign}\left(\textup{thr}_h-|\mathcal C_u^k|\right)+1,
\end{aligned}
\end{equation}
where $|\mathcal C_u^k|$ represents the size of $\mathcal{C}_u^k$, $\textup{sign}(\cdot)$ represents the signum function, $\textup{thr}_c$ \textcolor[rgb]{0.00,0.00,0.00}{and $\textup{thr}_h$ represent} the minimum size required for a connected domain to be considered as having a propagation path \textcolor[rgb]{0.00,0.00,0.00}{and} having a high-order NLOS path\textcolor[rgb]{0.00,0.00,0.00}{, respectively}. If size $|\mathcal C_u^k|$ of a connected domain is greater than $\textup{thr}_c$ \textcolor[rgb]{0.00,0.00,0.00}{but} less than $\textup{thr}_h$, then we set $c_u^k=2$, which implies the connected domain $\mathcal C_u^k$ encompassing a useful propagation path; otherwise, we set $c_u^k=0$, which implies the connected domain $\mathcal C_u^k$ not encompassing a profitable path. Moreover, if $\mathcal C_u^k$ encompasses a profitable path, then the angles of the highest received signal power in $\mathcal C_u^k$ are the elevation AoD, azimuth AoD, elevation AoA, and azimuth AoA of this path.



\begin{figure}
  \centering
  \includegraphics[width=0.35\textwidth]{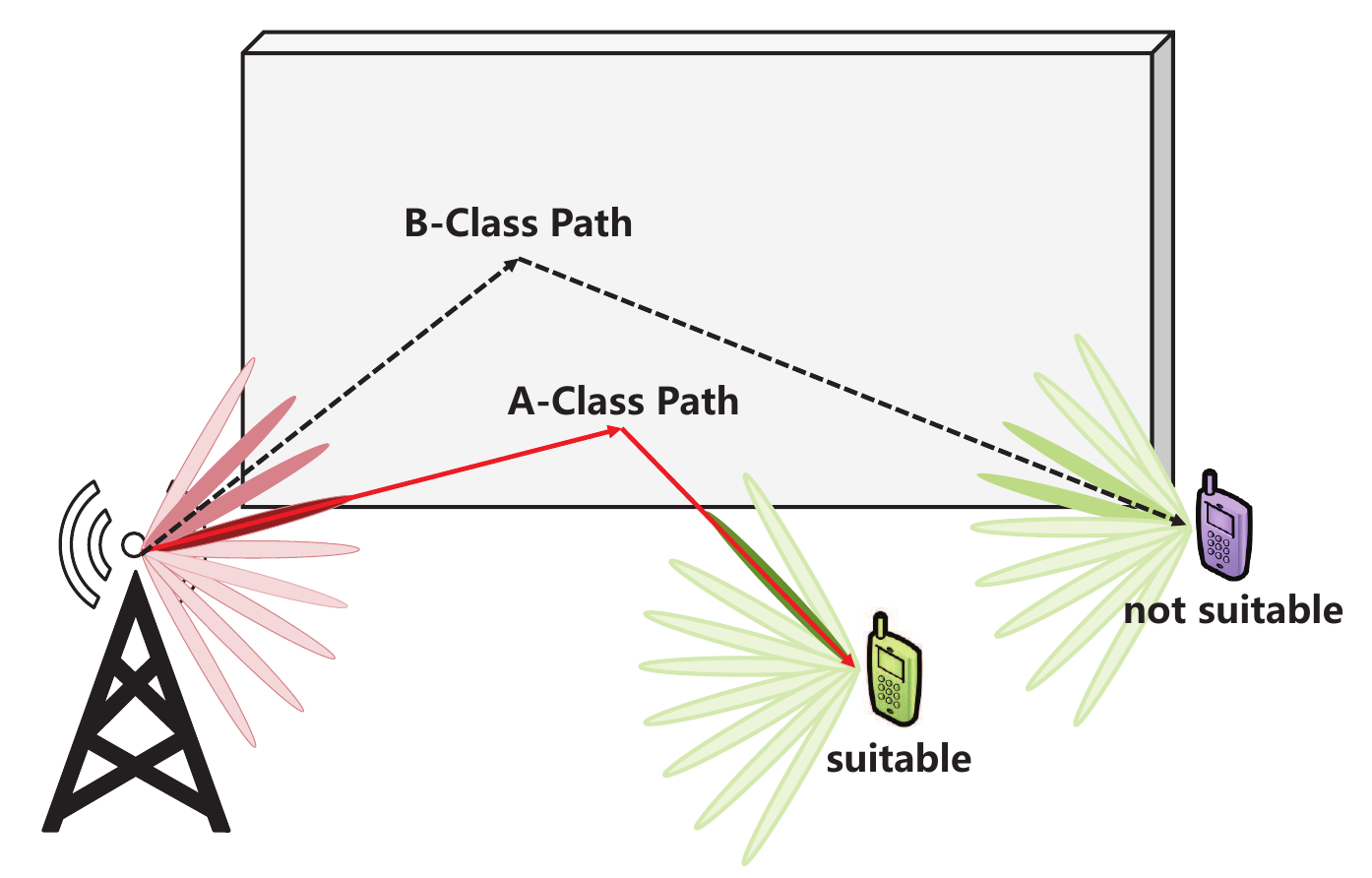}\\
  \caption{The AoA and AoD of the NLOS path may not align precisely with the angles in the codebook.}\label{Power_Factor}
\end{figure}

\subsection{Distinguish The LOS Path and The First-Order NLOS path}\label{s_reflection_factor}
After obtaining the connected domain with a profitable path, we need to recognize whether this path is a LOS path or a first-order NLOS path.
Denote $\theta_t^{u,k}$, $\phi_t^{u,k}$, $\theta_r^{u,k}$, and $\phi_r^{u,k}$ as the elevation AoD, azimuth AoD, elevation AoA, and azimuth AoA of the path in $\mathcal C_u^k$\textcolor[rgb]{0.00,0.00,0.00}{, respectively}.
Note that if the AoA and AoD of a path are complementary angles satisfying $\textup{tan}(\theta_t^{u,k})=\textup{tan}(\pi-\theta_r^{u,k})$ and $\textup{tan}(\phi_t^{u,k})=\textup{tan}(\pi-\phi_r^{u,k})$, then this path is deemed as a LOS path; otherwise, it is deemed as a first-order NLOS path.
Hence we design the reflection factor as
\begin{equation}\label{reflection_factor}
\begin{aligned}
r_u^k=\textup{sign}\left[\textup{thr}_{tan}-|\textup{tan}(\theta_t^{u,k})-\textup{tan}(\pi-\theta_r^{u,k})|\right.\\
\left.-|\textup{tan}(\phi_t^{u,k})-\textup{tan}(\pi-\phi_r^{u,k})|\right]+1,
\end{aligned}
\end{equation}
where $\textup{thr}_{tan}$ is the maximum LOS tolerance of the tangent values' difference between the AoA and AoD. In other words, if {$|\textup{tan}(\theta_t^{u,k})-\textup{tan}(\pi-\theta_r^{u,k})|+ |\textup{tan}(\phi_t^{u,k})-\textup{tan}(\pi-\phi_r^{u,k})| \leq \textup{thr}_{tan}$}, then we deem the path among $\mathcal C_u^k$ as a LOS path; otherwise, we deem this path as a first-order NLOS path.\footnote{The introduction of $\textup{thr}_{tan}$ aims to enhance the robustness of NLOS path detection. For instance, when the AoA and AoD of a LOS path are not centered on the grid, their estimations become imperfect, resulting in non-zero values of $|\textup{tan}(\theta_t^{u,k})-\textup{tan}(\pi-\theta_r^{u,k})|$ and $|\textup{tan}(\phi_t^{u,k})-\textup{tan}(\pi-\phi_r^{u,k})|$. Then the path will be mistakenly identified as an NLOS path, leading to an erroneous point calculated from this path.}

\subsection{Determine Whether \textcolor[rgb]{0.00,0.00,0.00}{A} Path Can Be Accurately Estimated}\label{s_power_factor}

We determine whether the angles of a path can be accurately estimated based on whether the angles of the path matches the angles in the codebook.
As shown in Fig. \ref{Power_Factor}, the AoA and AoD of a path may not align precisely with the angles in the codebook, resulting in angle estimation error.
We then define paths with AoA and AoD aligning precisely with the angles in the codebook as A-Class paths, and paths with either AoA or AoD not aligning precisely with the angles in the codebook as B-Class paths.
\textcolor[rgb]{0.00,0.00,0.00}{From Fig. \ref{Power_Factor},} the AoA and AoD estimations of A-Class paths are more accurate compared to those of B-Class paths.
Hence, we believe that a path can be accurately estimated only if it is an A-Class path.

Assume $qd_k^u=(i_k^u,j_k^u,p_k^u,q_k^u)$ is the coordinate of the angles with the highest power among $\mathcal C^u_k$, and denote the adjacent set of $qd_k^u$ as $\mathcal D^u_k$, i.e.
\begin{equation}\label{D_k_u}
\begin{aligned}
\mathcal D^u_k=\{(i_k^u-1,j_k^u,p_k^u,q_k^u),(i_k^u+1,j_k^u,p_k^u,q_k^u),\\
(i_k^u,j_k^u-1,p_k^u,q_k^u),(i_k^u,j_k^u+1,p_k^u,q_k^u),\\
(i_k^u,j_k^u,p_k^u-1,q_k^u),(i_k^u,j_k^u,p_k^u+1,q_k^u),\\
(i_k^u,j_k^u,p_k^u,q_k^u-1),(i_k^u,j_k^u,p_k^u,q_k^u+1)\}.
\end{aligned}
\end{equation}
Denote $\widehat{qd_k^u}$ and $\widecheck{qd_k^u}$ as the coordinates of the angles with the highest and the lowest powers among $\mathcal D^u_k$.
In the connected domain of an A-Class path, the highest power is obviously higher than the powers of adjacent angles. Consequently, the \textcolor[rgb]{0.00,0.00,0.00}{power difference} among the angles in the adjacent set \textcolor[rgb]{0.00,0.00,0.00}{is} minimal, leading to a small value of the ratio $\frac{\mathbf P_u[\widehat{qd}_k^u]}{\mathbf P_u[\widecheck{qd}_k^u]}$.
Conversely, in the connected domain of a B-Class path, there may be one or more angles in the adjacent set with considerably higher powers, resulting in a large value of the ratio $\frac{\mathbf P_u[\widehat{qd}_k^u]}{\mathbf P_u[\widecheck{qd}_k^u]}$.
Hence, we design a power factor to indicate the existence of the A-Class path as
\begin{equation}\label{power_factor}
\begin{aligned}
p_u^k=\textup{sign}\left[\textup{thr}_{pow}-\frac{\mathbf P_u[\widehat{qd}_k^u]}{\mathbf P_u[\widecheck{qd}_k^u]}\right]+1,
\end{aligned}
\end{equation}
where $\textup{thr}_{pow}$ refers to the maximum value of $\frac{\mathbf P_u[\widehat{qd}_k^u]}{\mathbf P_u[\widecheck{qd}_k^u]}$ for a connected domain to be considered as having an A-Class path.
For multi-user environment reconstruction, excluding the B-Class paths may make point set $\mathcal P$ sparser but more accurate.
Conversely, incorporating the $\text{B-Class}$ paths will bring numerous unacceptable error points.
Therefore, we only utilize the A-Class paths to calculate the reflection points.

Based on the connectivity factor, the reflection factor, and the power factor, we can calculate the user selection factor as
\begin{equation}\label{criterion}
\begin{aligned}
s_u&=(\sum_{k=1}^{n_u}s_{u,los}^{k})\cdot(\sum_{k=1}^{n_u}s_{u,nlos}^{k})\\
&=\left[\sum_{k=1}^{n_u}\frac{1}{8}c_u^k\cdot (2-r_u^k)\cdot p_u^k \right]\cdot \left[\sum_{k=1}^{n_u}\frac{1}{8}c_u^k\cdot r_u^k\cdot p_u^k\right],
\end{aligned}
\end{equation}
where $\sum_{k=1}^{n_u}\frac{1}{8}c_u^k\cdot (2-r_u^k)\cdot p_u^k$ and $\sum_{k=1}^{n_u}\frac{1}{8}c_u^k\cdot r_u^k\cdot p_u^k$ represent the number of LOS paths and NLOS paths that can be accurately calculated from $\mathbf P_u$.
Moreover, $s_u$ represents the number of reflection points that can be accurately calculated from $\mathbf P_u$.
Hence, $s_u\geq 1$ indicates that the user $u$ is a superior user; \textcolor[rgb]{0.00,0.00,0.00}{otherwise} $s_u = 0$.

\subsection{Calculate The Reflection Points}
For a superior user, we first use OFDM ranging to calculate the length of the LOS path \cite{sun2022indoor,10271123}.

In an OFDM system, signals are modulated onto subcarriers with different frequencies to facilitate transmission. For the same distance, the received signals of \textcolor[rgb]{0.00,0.00,0.00}{different} subcarriers exhibit \textcolor[rgb]{0.00,0.00,0.00}{different phase shifts}.
During the beam sweeping stage, the received phase of the $m$-th subcarrier is $\frac{2\pi f_m d}{c}$ \textcolor[rgb]{0.00,0.00,0.00}{when beamforming is performed on the LOS direction}, where $d$ is the length of the LOS path, $f_m$ is the frequency of the $m$-th subcarrier, and $c$ is the speed of light. Hence, the length of the LOS path can be estimated by    
\begin{equation}\label{distace_estimation}
\begin{aligned}
d^{\star}=\underset{d}{\arg \max }\left|\sum_{m=1}^M \exp \left(\mathrm{j}\left(\varphi_m-\frac{2 \pi f_m d}{c}\right)\right)\right|,
\end{aligned}
\end{equation}
where $\varphi_m$ is the phase of the $m$-th subcarrier measured at the receiver.
Denote the location of the BS as $(x_b,y_b,z_b)$, the elevation AoA of the LOS path as $\theta_{los}$, and the azimuth AoA of the LOS path as $\phi_{los}$. Then, location $(x_u,y_u,z_u)$ of the user can be {calculated~by}
\begin{equation}\label{location}
\begin{aligned}
\left\{\begin{aligned}
&\frac{y_u-y_b}{x_u-x_b}=\tan (\phi_{los}), \\
&\frac{z_u-z_b}{\sqrt{\left(x_u-x_b\right)^2+\left(y_u-y_b\right)^2}}=\tan \left(\frac{\pi}{2}-\theta_{los}\right), \\
&\left\|\left[x_u, y_u, z_u\right]^T-\left[x_b, y_b, z_b\right]^T\right\|_2+\\
&\left\|\left[x_u, y_u, z_u\right]^T-\left[x_b, y_b, z_b\right]^T\right\|_2=d^{\star}.
\end{aligned}\right.
\end{aligned}
\end{equation}

After obtaining the location of the user, we utilize azimuth AoD $\phi_{t,nlos}$, elevation AoD $\theta_{t,nlos}$, azimuth AoA $\phi_{r,nlos}$, and elevation AoA $\theta_{r,nlos}$ of the first-order NLOS path to calculate the reflection point $(x_u^p,y_u^p,z_u^p)$ as
\begin{equation}\label{reflection_point}
\begin{aligned}
\left\{\begin{aligned}
&\frac{y_u^p-y_b}{x_u^p-x_b}=\tan (\phi_{t,nlos}), \\
&\frac{z_u^p-z_b}{\sqrt{\left(x_u^p-x_b\right)^2+\left(y_u^p-y_b\right)^2}}=\tan \left(\frac{\pi}{2}-\theta_{t,nlos}\right), \\
&\frac{y_u^p-y_u}{x_u^p-x_u}=\tan (\phi_{r,nlos}), \\
&\frac{z_u^p-z_u}{\sqrt{\left(x_u^p-x_u\right)^2+\left(y_u^p-y_u\right)^2}}=\tan \left(\frac{\pi}{2}-\theta_{r,nlos}\right).
\end{aligned}\right.
\end{aligned}
\end{equation}
By recording the reflection points of the superior users, we obtain the point set $\mathcal P$.

\subsection{Surface Fitting from The Points}
However, the point set is a discrete approximation of environment reconstruction.
We then smoothen the environment reconstruction result by generating statistical surfaces that approximate the points in $\mathcal P$.
Specifically, we use the $\text{K-Means}$ algorithm \textcolor[rgb]{0.00,0.00,0.00}{in~\cite{ahmed2020k}} to cluster the points in $\mathcal P$ and then fit a surface to the points in each cluster.
We propose to represent each surface by a high-order polynomial,
\begin{equation}\label{polynomial}
\begin{aligned}
z=&c_0 + c_1x + c_2y+c_3x^2+c_4xy+c_5y^2+\\
&c_6x^3+c_7x^2y+c_8xy^2+c_9y^3,
\end{aligned}
\end{equation}
where $c_0,c_1,\cdots,c_9$ are the coefficients of the polynomial and $(x,y,z)$ is the coordinate of the point in the surface. Next, we calculate $c_0,c_1,\cdots,c_9$ by the following steps.
\begin{itemize}
\item[$\bullet$] Denote $z_1(x,y) = c_0+c_1x+c_2y$. We employ multivariate linear regression (MLR) \cite{draperwiley} to fit the plane $z_1$ and obtain proper $c_0$, $c_1$, and $c_3$ that ensures a high degree of proximity between the points in $\mathcal P$ and the plane $z_1(x,y)$.
\item[$\bullet$] Denote $z_2(x,y) = c_3w_1+c_4w_2+c_5w_3+z_1(x,y)$, where $w_1=x^2$, $w_2=xy$, and $w_3=y^2$. We utilize MLR to fit $z_2(x,y)$ and obtain proper $c_3$, $c_4$, and $c_5$.
\item[$\bullet$] Denote $z(x,y) = c_6w_4+c_7w_5+c_8w_6+c_9w_7+z_1(x,y)$, where $w_4=x^3$, $w_5=x^2y$, $w_6=xy^2$, and $w_7=y^3$. We utilize MLR to fit $z(x,y)$ and obtain proper $c_6$, $c_7$, $c_8$, and $c_9$.
\end{itemize}

\section{Multi-Modal Fusion Based Environment Reconstruction}\label{s_environment_reconstruction}
\begin{figure*}[t]
  \centering
  \includegraphics[width=0.75\textwidth]{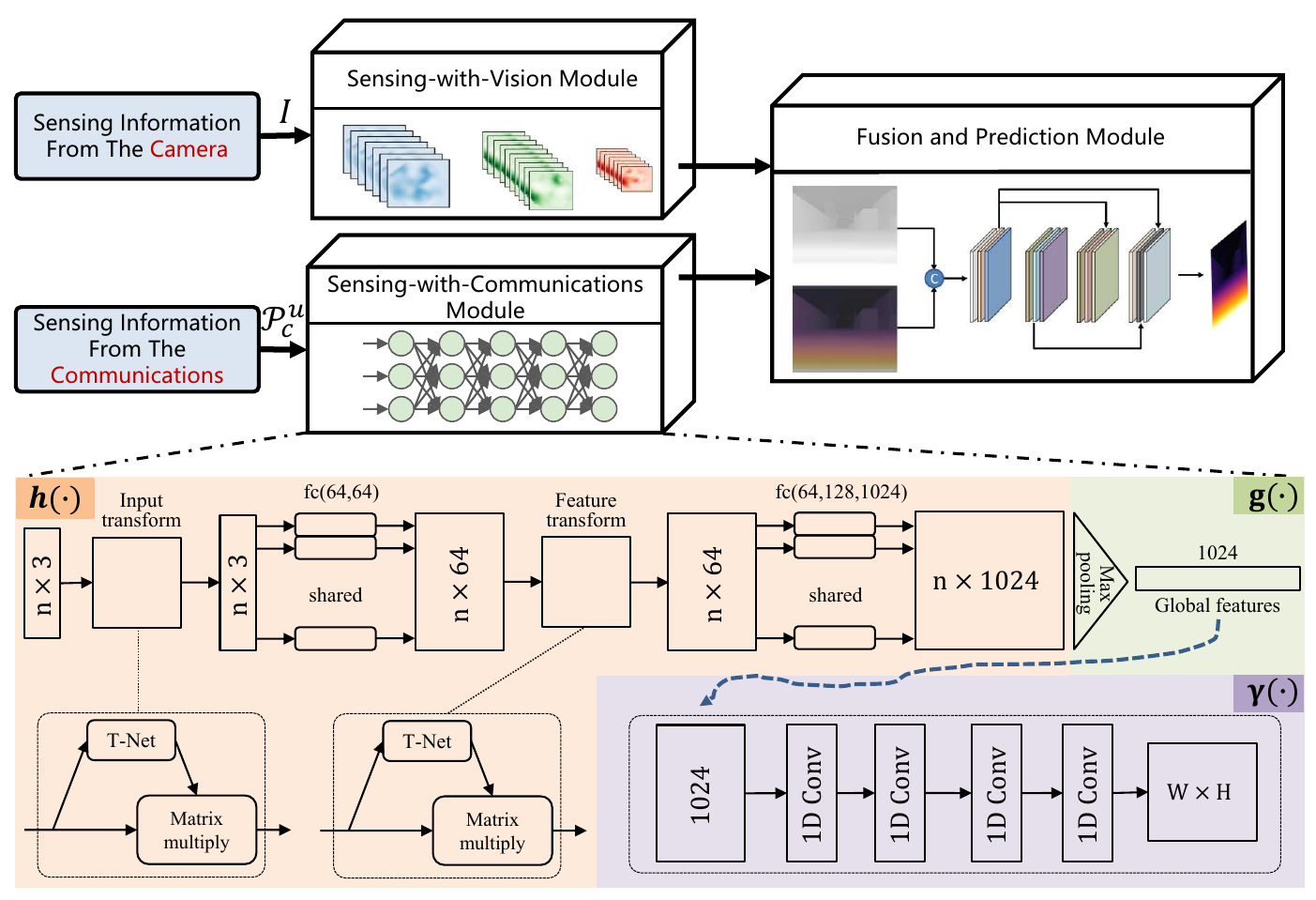}\\
  \caption{The structure of the multi-modal fusion network.}\label{MMFN}
\end{figure*}

Although the environment reconstruction based on multi-user communications has denser points than that based on single-user communications, it still cannot support complex sensing tasks, such as autonomous driving.
Assume that the users are equipped with both communications devices and cameras.
We \textcolor[rgb]{0.00,0.00,0.00}{will} leverage both the vision sensing and ISAC to obtain more comprehensive environment reconstruction.
\textcolor[rgb]{0.00,0.00,0.00}{As shown in Fig.~\ref{MMFN},} the user first downloads the point set $\mathcal P$ from the BS, and then resorts to a multi-modal fusion network (MMFN) to fuse $\mathcal P$ and the image for depth estimation.
The MMFN consists of a sensing-with-vision (SWV) module to extract the features from the image, a sensing-with-communications (SWC) module to extract the features from point set $\mathcal P$, and a fusion \textcolor[rgb]{0.00,0.00,0.00}{and} prediction (FP) module to fuse the features and predict the depth map.

\subsection{Sensing-with-Vision Module}
We adopt the state-of-the-art Mixing Datasets for Zero-shot Cross-dataset Transfer (MiDaS) \cite{ranftl2020towards} as the SWV module, which has been pretrained on 10 distinct datasets to ensure high quality and great generalization.

\subsection{Sensing-with-Communications Module}
Note that, point set $\mathcal P$ provided by the BS is based on the world coordinate system while the image is based on the camera coordinate system \cite{tsai1987versatile}.
In order to achieve better fusion results, the point set $\mathcal P$ and the image should be in the same coordinate system.
Hence, we first convert $\mathcal P$ to the camera coordinate system.

Denote the rotation angles of the camera along each axis as $(\phi_x,\phi_y,\phi_z)$.
Then the rotation matrices along each axis are
\begin{equation}\label{rotation-matrices}
\begin{aligned}
\mathbf{R}_x =& \left[
\begin{matrix}
\textup{cos}\phi_x & -\textup{sin}\phi_x & 0 \\
\textup{sin}\phi_x & \textup{cos}\phi_x & 0 \\
0 & 0 & 1
\end{matrix}
\right],\\
\mathbf{R}_y = &\left[
\begin{matrix}
1 & 0 & 0 \\
0 & \textup{cos}\phi_y & \textup{sin}\phi_y \\
0 & -\textup{sin}\phi_y & \textup{cos}\phi_y
\end{matrix}
\right],\\
\mathbf{R}_z = &\left[
\begin{matrix}
\textup{cos}\phi_z & 0 & -\textup{sin}\phi_z \\
0 & 1 & 0 \\
\textup{sin}\phi_z & 0 & \textup{cos}\phi_z
\end{matrix}
\right].
\end{aligned}
\end{equation}
The rotation matrix of the camera can be calculated by $\mathbf{R} = \mathbf{R}_x\cdot \mathbf{R}_y\cdot \mathbf{R}_z$.

Denote the homogeneous coordinate\footnote{Homogeneous coordinates use N+1 dimensions to represent N-dimensional coordinates to deal with geometric problems in perspective space \cite{maxwell1952methods}. In perspective space, two parallel lines can meet at infinity. Using homogeneous coordinates, the translation of an object can be conveniently represented by a linear transformation.} of the reflection point in the world coordinate system as $(x_u^p,y_u^p,z_u^p,1)$, and denote the relative displacement of the camera and the user as $\mathbf{T}=[t_x,t_y,t_z]^T$.
Then the coordinate of the point in the camera coordinate system \textcolor[rgb]{0.00,0.00,0.00}{will be}
\begin{equation}\label{camera-coordinate}
\left[
\begin{matrix}
x_{u,c}^p \\ y_{u,c}^p \\ z_{u,c}^p \\ 1
\end{matrix}
\right]=
\left[
\begin{matrix}
\mathbf{R} & \mathbf{T} \\
0 & 1
\end{matrix}
\right]
\left[
\begin{matrix}
x_u^p \\ y_u^p \\ z_u^p \\ 1
\end{matrix}
\right].
\end{equation}

After converting all points in $\mathcal P$ to the camera coordinate system, we record them in a new set $\mathcal P_c^u=\left\{\bm x_1, \ldots, \bm x_n\right\}$ and input $\mathcal P_c^u$ into the SWV module.
However, the point set $\mathcal P_c^u$ is unordered, which means that changing the order of points in $\mathcal P_c^u$ does not alter the information it contains.
Hence, we introduce specific symmetrizations to ensure that the output features of the SWV module remain consistent regardless of the input points' order.
The symmetrical function of the SWC module is designed as
\begin{equation}\label{input_transform}
\begin{aligned}
f\left(\left\{\bm x_1, \ldots, \bm x_n\right\}\right) = \gamma( g\left(h\left(\bm x_1\right), \ldots, h\left(\bm x_n\right)\right)),
\end{aligned}
\end{equation}
where $h(\cdot)$ represents the function of the multi-layer perceptron (MLP) shared by all points, $g(\cdot)$ represents the max pooling function, which is a symmetric function, and $\gamma(\cdot)$ represents the function of the MLP for feature extraction.

The SWC module is shown in Fig. \ref{MMFN}.
In order to enhance the network's adaptability to different input point sets, we let the points in $\mathcal P_c^u$ undergo an input transform process\cite{Qi_2017_CVPR}, which mainly relies on a T-net.
The T-net generates an affine transformation matrix, which is then applied directly to the input points. Next, a shared  fully-connected (FC) layer is employed to extract features for each individual point. Following the FC, a ``feature transform'' module is utilized to transform the extracted features into a suitable domain using another T-net. The transformed features are subsequently passed through an FC to obtain deep features. The deep features are passed into a max pooling function to generate the global features. Then the global features are fed into four 1D CNN layers.
The output of the final CNN is reshaped to $\bm F_{cs}\in \mathbb{R}^{W\times H}$.

\subsection{Fusion \textcolor[rgb]{0.00,0.00,0.00}{and} Prediction Module}
The structure of the FP module is shown in Fig. \ref{MMFN}.
Denote the output the SWV module as $\bm D_p\in \mathbb R^{W\times H}$.
We combine $\bm D_p$ and $\bm F_{cs}$ as $\bm F = \text{cat}(\bm F_{cs},\bm D_p)$.
Then $\bm F$ is input into several CNNs with skip connections.
The final output of the fusion module is the predicted depth map~$\hat{\bm D}$.

We utilize the root mean-squared error (RMSE) of the predicted depth for all pixels as the loss function
\begin{equation}\label{loss_func}
\begin{aligned}
\mathcal L=\frac{1}{N}\sqrt{\frac{1}{W\times H} \sum_{d_i\in \bm D, \hat d_i \in \hat{\bm D}}\left|d_i-\hat d_i\right|^2},
\end{aligned}
\end{equation}
where $\bm D$ is the ground truth of the depth map and $N$ is the size of the dataset.

\subsection{Meta-Learning \textcolor[rgb]{0.00,0.00,0.00}{based} Training Strategy for Any User Quantity}
For MMFN, if the quantity of input points changes, then we have to train a new network; otherwise, the performance of MMFN may suffer significant degradation. To \textcolor[rgb]{0.00,0.00,0.00}{generalize the} network across different user quantities, we design a meta-learning-based\footnote{Meta-learning \cite{finn2017model} is a machine learning approach that focuses on developing algorithms or models capable of learning and adapting to new tasks or environments quickly.
The training methodology of meta-learning involves training a model for a specific number of epochs on a given task and utilizing the parameters of that model as the initial values for training another task.} training strategy.
Specifically, we denote the depth estimation through the fusion of images and $N_m^u$ users' points as task-$m$, where $N_m^u\in \{N_1^u,N_2^u,\cdots,N_M^u\}$ is the possible user quantity. Then we train MMFN for these tasks sequentially, where the initial parameters of task-$(m+1)$ is the trained parameters {of~$\text{task-}m$}.
\begin{algorithm}[t]
\caption{Meta Learning Based Training Strategy for Depth Estimation}
\label{alg1}
\begin{algorithmic}
\REQUIRE Training dataset $\mathcal{D}$, number of iterations $n_{iter}$, number of iterations of each task $n_m$, initialized trainable parameters of the SWC module $\bm{\Theta}_c$, initialized trainable parameters of the FP module $\bm{\Theta}_f$, trained parameters of MiDaS $\bm{\Theta}_M$, and the maximum number of users $N_{max}$.
\ENSURE Trained parameters of SWC module $\bm{\Theta}_c$ and trained parameters of the FP module $\bm{\Theta}_f$.
\FOR{k = 1 to $n_{iter}$}
\FOR{n = 1 to $n_{m}$}
\STATE - Draw mini-batch $\mathcal{D}_{k}$: a random subset of $\mathcal{D}_m$
\STATE - Prepare the input of MiDaS $\bm I$
\STATE - Generate the input of the SWC module $\bm P_c^m$ by masking $(N_max-m)$ rows
\STATE - Estimate the dense depth $\bm D_i$ by MiDaS and the image~$\bm I$
\STATE - Estimate the domain map $\bm F_{dm}$ by the SWC module and the point cloud $\bm P_c^m$
\STATE - Calculate the output depth by fusing $\bm D_i$ and $\bm F_{dm}$ based on fusion module
\STATE - Compute the loss: $\mathcal{L}$
\STATE Back-propagation Phase:
\STATE - Use Adam optimizer to update $\bm{\Theta_{c}}$ and $\bm{\Theta_{f}}$
\ENDFOR
\ENDFOR
\end{algorithmic}
\end{algorithm}
\begin{figure}[t]
  \centering
  \includegraphics[width=0.15\textwidth]{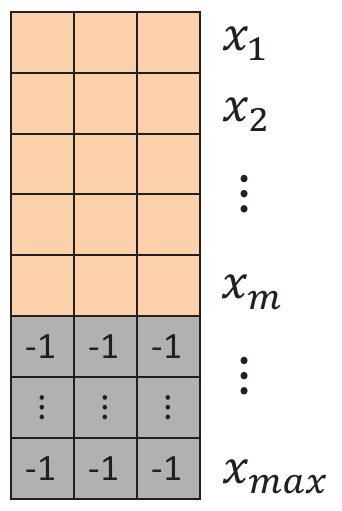}\\
  \caption{The mask method for the input of the communications sensing module.}\label{masking}
\end{figure}


However, when the quantity of input points varies, the input dimension of the communications-sensing module also changes. We then design an input masking method illustrated in Fig. \ref{masking} to ensure the immutability of the input dimension. Assume that the maximum number of users is $N_{max}$. Then we generate a matrix $\bm P_{m}\in R^{N_{max}\times 3}$ as the input for task-$m$ whose first $m$ rows is the $m$ points' coordinates in the camera coordinate system and the last $(N_{max}-m)$ rows are masked as $-1$.
The meta-learning based training strategy is illustrated in Algorithm~\ref{alg1}.


\begin{table*}[t]
\renewcommand\arraystretch{}
\caption{Environment Reconstruction under Different User Selection Factors}\label{table1}
\centering
\begin{tabular}{m{1cm}<{\centering}|m{1.7cm}<{\centering}|m{1.7cm}<{\centering}|m{1.8cm}<{\centering}|m{1.3cm}<{\centering}|m{1.8cm}<{\centering}|m{1.8cm}<{\centering}|m{1.8cm}<{\centering}|m{1cm}<{\centering}}
\hline
\hline
Factor & Without Selection & Connectivity & Reflection & Power & Connectivity + Reflection & Connectivity + Power & Reflection + Power & ALL \\
\hline
RMSE & $0.9233$ & $0.8535$ & $0.5141$ & $0.4351$ & $0.2266$ & $0.1584$ & $0.4026$ & $\bf{0.0556}$  \\
\hline
\hline
\end{tabular}
\end{table*}

\subsection{Evaluation Metrics of Depth Estimation}

The evaluation metrics of depth estimation includes \textcolor[rgb]{0.00,0.00,0.00}{the following.}
\begin{itemize}
\item[$\bullet$] Root \textcolor[rgb]{0.00,0.00,0.00}{Mean-Squared} Error (RMSE):
\begin{equation*}
\text{RMSE}=\sqrt{\frac{1}{N} \sum_{d_i\in \bm D, \hat d_i \in \hat{\bm D}}\left|d_i-\hat{d}_i\right|^2}\textcolor[rgb]{0.00,0.00,0.00}{.}
\end{equation*}

\item[$\bullet$] Root \textcolor[rgb]{0.00,0.00,0.00}{Mean-Squared} Error logscale (RMSElog):
\begin{equation*}
\text{RMSElog}=\sqrt{\frac{1}{N} \sum_{d_i\in \bm D, \hat d_i \in \hat{\bm D}}\left|\log d_i-\log \hat{d}_i\right|^2}.
\end{equation*}

\item[$\bullet$] Mean-Absolute Error (MAE):
\begin{equation*}
\text{MAE}=\sqrt{\frac{1}{N} \sum_{d_i\in \bm D, \hat d_i \in \hat{\bm D}}\left|d_i-\hat{d}_i\right|}.
\end{equation*}

\item[$\bullet$] Root \textcolor[rgb]{0.00,0.00,0.00}{Mean-Squared} Error logscale (MAElog):
\begin{equation*}
\text{MAElog}=\sqrt{\frac{1}{N} \sum_{d_i\in \bm D, \hat d_i \in \hat{\bm D}}\left|\log d_i-\log \hat{d}_i\right|}.
\end{equation*}

\item[$\bullet$] Absolute Relative Error (AbsRel):
\begin{equation*}
\text{AbsRel}=\frac{1}{N} \sum_{d_i\in \bm D, \hat d_i \in \hat{\bm D}} \frac{\left|d_i-\hat{d}_i\right|}{d_i}.
\end{equation*}

\item[$\bullet$] Square-Relative Error (SqRel):
\begin{equation*}
\text{SqRel}=\frac{1}{N} \sum_{d_i\in \bm D, \hat d_i \in \hat{\bm D}} \frac{\left|d_i-\hat{d}_i\right|^2}{d_i}.
\end{equation*}

\item[$\bullet$] $\delta_n$ threshold:
\begin{equation*}
\delta_n=\left|\left\{\hat{d}_i: \max_{d_i\in \bm D, \hat d_i \in \hat{\bm D}} \left(\frac{\hat{d}_i}{d_i}, \frac{d_i}{\hat{d}_i}\right)<1.25^n\right\}\right|/|\bm D|.
\end{equation*}
\end{itemize}

\section{Simulation Results}\label{simulation}
In this section, we generate the dataset and evaluate the performance of the proposed multi-user selection and multi-modal fusion based environment reconstruction.
\subsection{Dataset Generation}
\begin{figure}[t]
  \centering
  \includegraphics[width=0.38\textwidth]{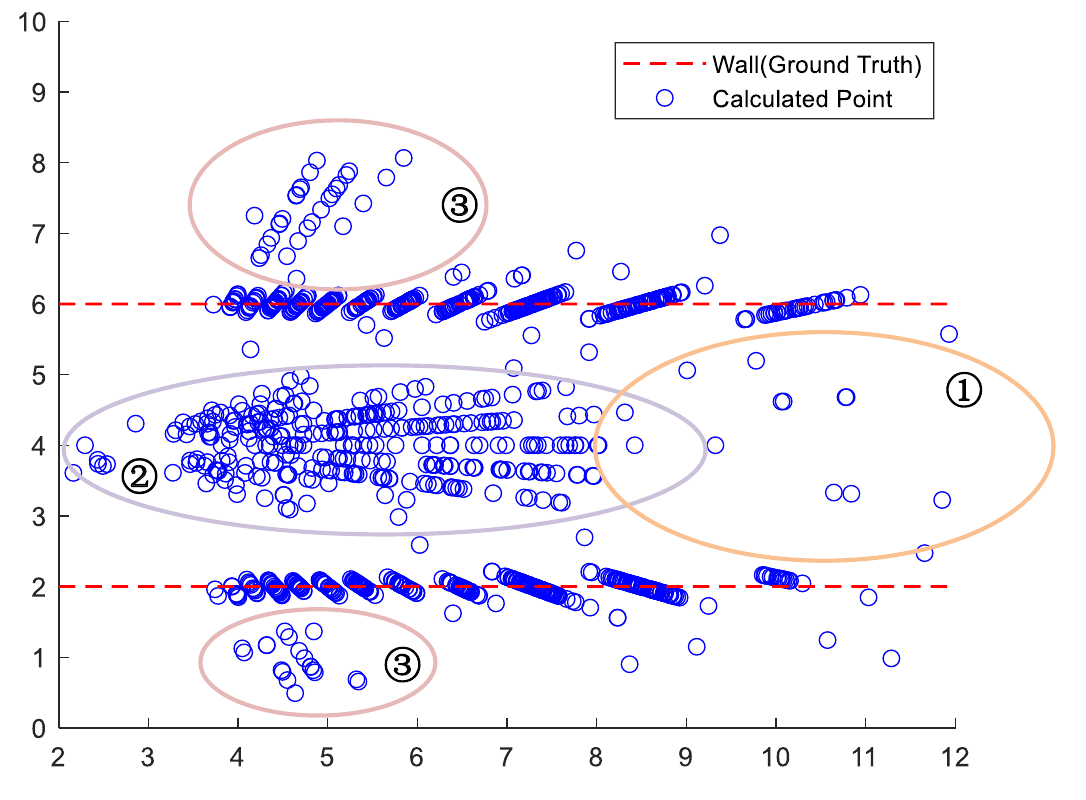}\\
  \caption{The calculated points without sensing user selection}\label{no_choose}
\end{figure}

\begin{figure}[t]
  \centering
  \includegraphics[width=0.38\textwidth]{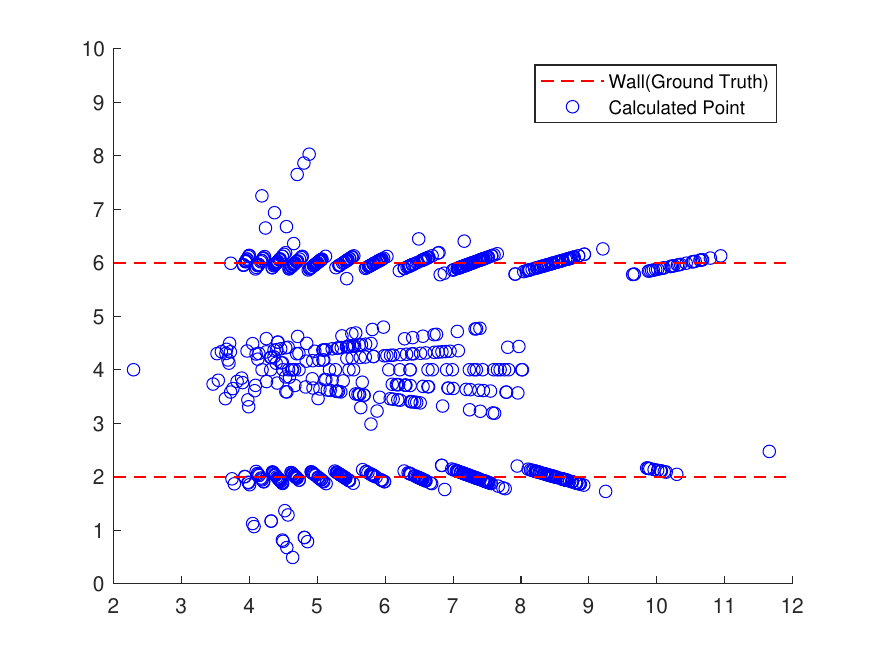}\\
  \caption{The environment reconstruction based on the connectivity factor.}\label{liantong}
\end{figure}

\begin{figure}[t]
  \centering
  \includegraphics[width=0.38\textwidth]{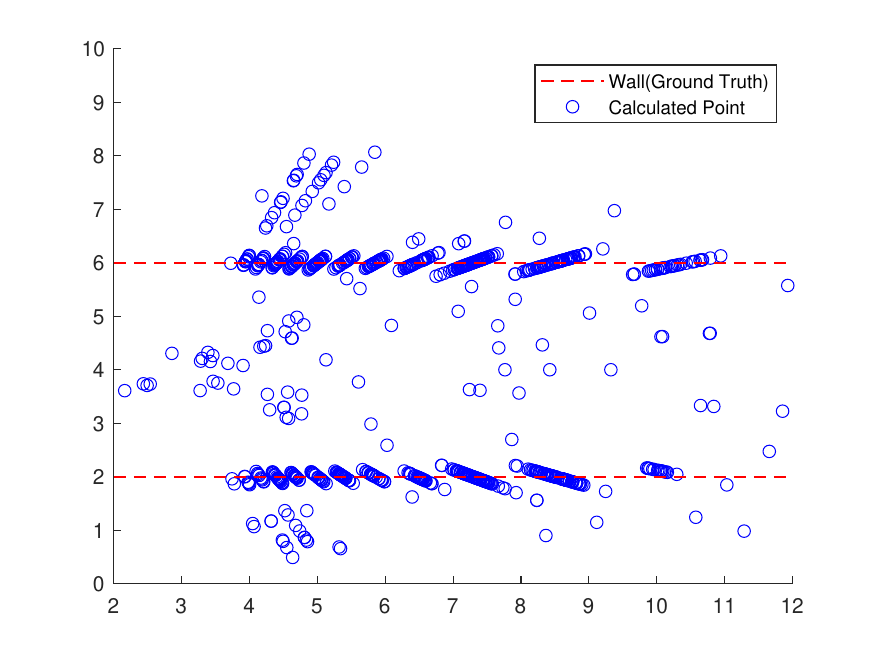}\\
  \caption{The environment reconstruction based on the reflection factor.}\label{tan}
\end{figure}

\begin{figure}[t]
  \centering
  \includegraphics[width=0.38\textwidth]{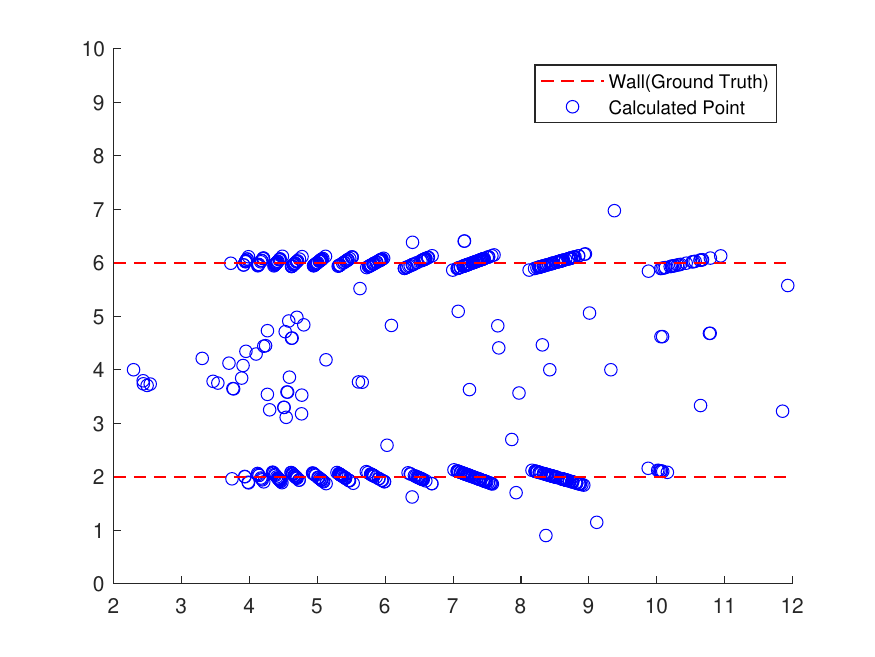}\\
  \caption{The environment reconstruction based on the power factor.}\label{power}
\end{figure}

\begin{figure}[t]
  \centering
  \includegraphics[width=0.4\textwidth]{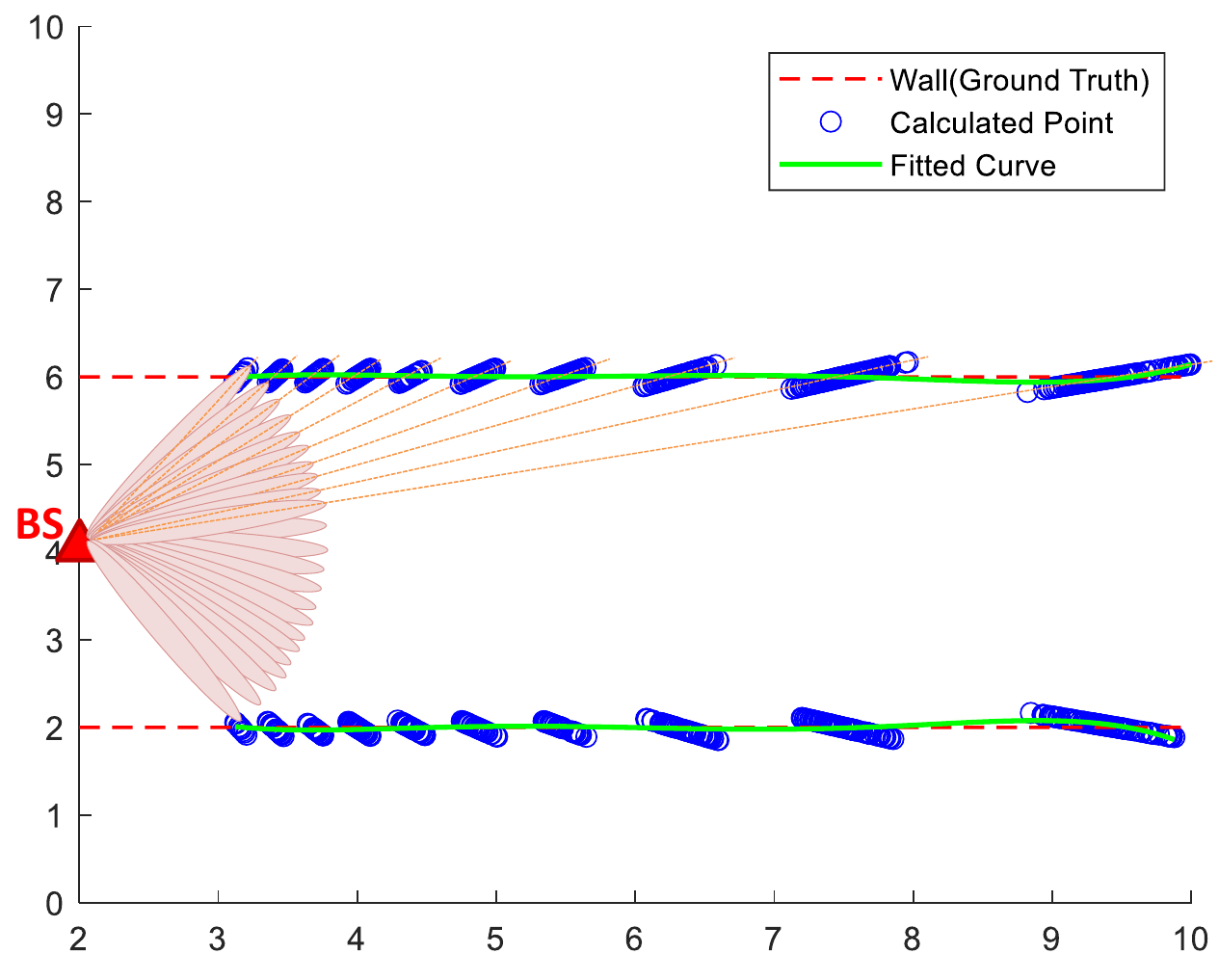}\\
  \caption{The environment reconstruction based on the proposed user selection algorithm.}\label{proposed}
\end{figure}

We consider a communications scenario involving numerous pedestrians equipped with communications devices as well as cars equipped with both communications devices and cameras.
We use CARLA \cite{dosovitskiy2017carla} to build the simulation scenario and utilize the cameras in the scenario to capture the images.\footnote{CARLA has been meticulously designed to streamline the development, training, and validation of autonomous driving systems. In addition to providing open-source code and protocols, CARLA offers a wealth of freely accessible open digital assets, including urban layouts, buildings, and vehicles, all tailored to this domain. CARLA empowers users with the ability to customize sensor suites and environmental conditions to suit their specific needs, while also enabling full control over both static and dynamic actors. Furthermore, CARLA boasts functionalities such as map generation, facilitating extensive experimentation and thorough testing of autonomous driving systems.}
Next, we use the 3D ray-tracing package ``propagationModel'' of MATLAB \cite{matlab} to calculate the wireless parameters. 
The communication frequency is set to 28 GHz and the bandwidth is set to 40 MHz. The BS is equipped with a uniform planar array (UPA) of $8\times 8$ antennas and the user is equipped with a UPA of $4\times 4$ antennas.

\begin{figure}[t]
  \centering
  \includegraphics[width=0.37\textwidth]{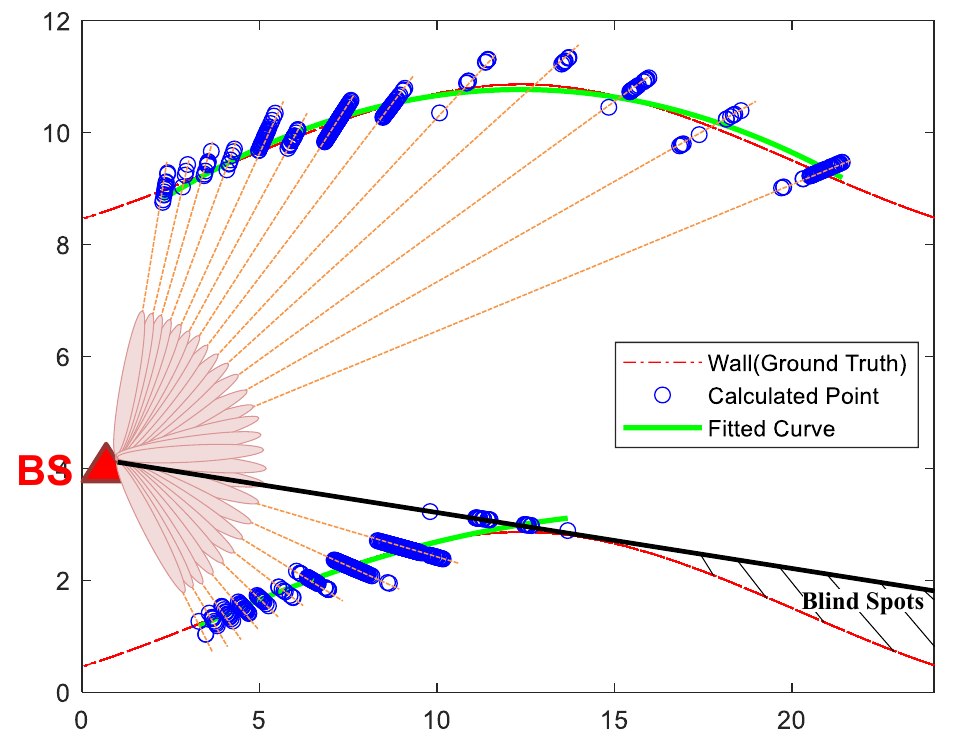}\\
  \caption{The environment reconstruction of irregular walls based on the proposed user selection algorithm.}\label{humian}
\end{figure}

\begin{figure}[t]
  \centering
  \includegraphics[width=0.43\textwidth]{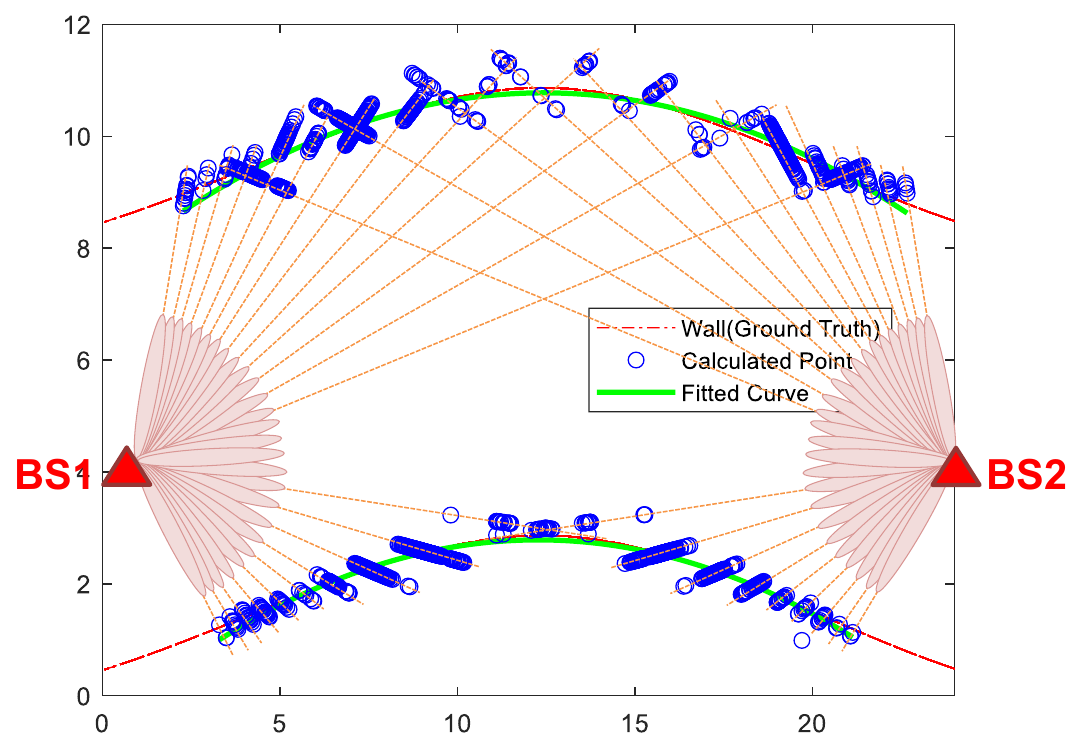}\\
  \caption{The environment reconstruction of irregular walls based on two BSs.}\label{humian_two_bs}
\end{figure}

\begin{figure}[t]
  \centering
  \includegraphics[width=0.41\textwidth]{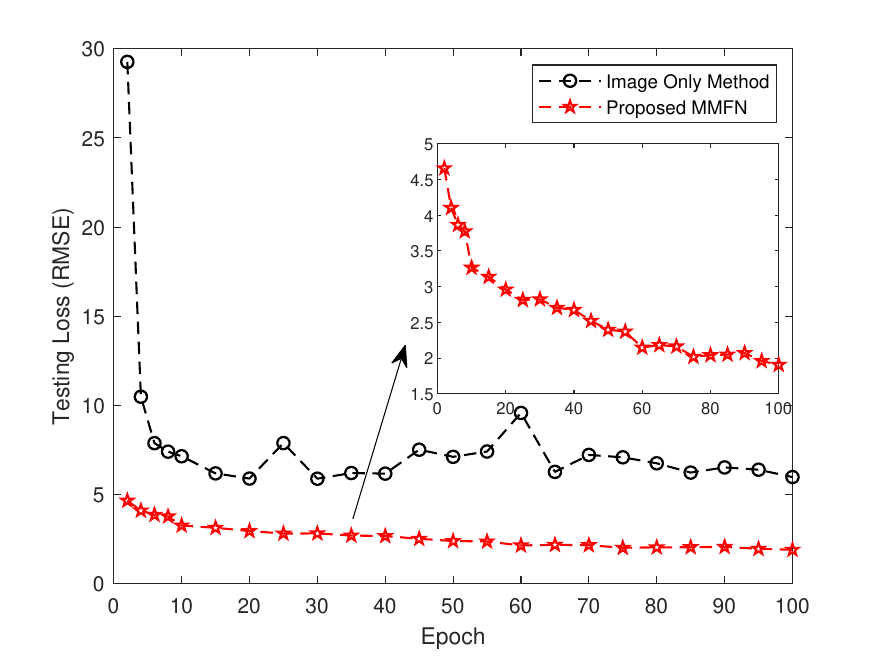}\\
  \caption{Testing loss of the proposed MMFN and the image only method.}\label{trainloss}
\end{figure}

\begin{figure}[t]
  \centering
  \includegraphics[width=0.35\textwidth]{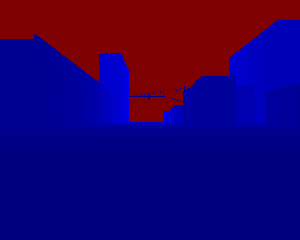}\\
  \caption{The ground truth of the depth map.}\label{gt}
\end{figure}

\begin{table*}[t]
\renewcommand\arraystretch{}
\caption{Depth estimation metrics under different weather conditions}\label{table2}
\centering
\begin{tabular}{m{1cm}<{\centering}|m{1.2cm}<{\centering}|m{1.3cm}<{\centering}|m{1.5cm}<{\centering}|m{1.2cm}<{\centering}|m{1.3cm}<{\centering}|m{1.4cm}<{\centering}|m{1.3cm}<{\centering}|m{1cm}<{\centering}|m{1cm}<{\centering}|m{1cm}<{\centering}}
\hline
\hline
Weather & Method & RMSE $\downarrow$ & $\text{RMSE}_{log}\downarrow$ & MAE $\downarrow$ & $\text{MAE}_{log}\downarrow$ & AbsRel $\downarrow$ & SqRel $\downarrow$ & $ \delta_1\uparrow $ & $ \delta_2\uparrow $ & $ \delta_3\uparrow $ \\
\hline
\multirow{2}*{Sunny} & image & 5.281 & 0.226 & 1.289 & 0.413 & 0.173 & 12.828 & 0.692 & 0.960 & 0.996 \\
\cline{2-11}
& fusion & \textbf{1.874} & 0.091 & 0.895 & 0.243 & 0.062 & 0.187 & 0.976 & 0.996 & 0.998 \\
\hline
\multirow{2}*{Rainy} & image & 10.315 & 0.437 & 2.054 & 0.538 & 0.281 & 15.557 & 0.554 & 0.815 & 0.906 \\
\cline{2-11}
& fusion & \textbf{3.564} & 0.176 & 1.355 & 0.362 & 0.126 & 0.524 & 0.812 & 0.977 & 0.994 \\
\hline
\multirow{2}*{Snowy} & image & 18.339 & 0.981 & 3.160 & 0.892 & 0.675 & 26.220 & 0.150 & 0.337 & 0.497 \\
\cline{2-11}
& fusion & \textbf{4.802} & 0.230 & 1.508 & 0.403 & 0.178 & 1.244 & 0.751 & 0.937 & 0.978 \\
\hline
\hline
\end{tabular}
\end{table*}

Without user selection, the points are calculated and presented in Fig. \ref{no_choose}, which are plagued by numerous errors.
Then we test the impact of the connectivity factor, the reflection factor, and the power factor on environment reconstruction. By separately incorporating these three factors, the environment reconstruction results are shown in Fig. \ref{liantong}, Fig. \ref{tan}, and Fig. \ref{power} respectively.
Note that the goal of the connectivity factor is to eliminate noise and high-order NLOS paths.
From Fig.~\ref{liantong}, many irregular noise points in circle \ding{172} of Fig.~\ref{no_choose} can be eliminated \textcolor[rgb]{0.00,0.00,0.00}{after introducing the connectivity factor}.
The purpose of the reflection factor is to avoid misidentifying the LOS path as a first-order NLOS path. 
From Fig. \ref{tan}, a large number of erroneous points between the BS and the user as shown in circle \ding{173} of Fig.~\ref{no_choose} can be eliminated \textcolor[rgb]{0.00,0.00,0.00}{after introducing the reflection factor}.
The objective of the power factor is to ensure that the selected user has an A-class path, which can potentially result in highly accurate environment reconstruction. As depicted in Fig.~\ref{power}, the power factor effectively eliminates numerous error points near the ground truth points in circle \ding{174} of Fig.~\ref{no_choose}.

The environment reconstruction result based on all three factors is shown in Fig.~\ref{proposed}.
An interesting phenomenon is that the calculated points all fall in the directions of the transmit beams and are clustered in distribution.
This is because that the points lie between these two clusters belong to the first-order NLOS paths whose angles do not align with the angles of the transmit beams, and the user selection algorithm excludes the first-order NLOS paths.
The RMSE under different user selection factors is shown in Tabel~\ref{table1}.
It can be seen that the RMSE of environment reconstruction without user selection is $0.9233$. However, by employing the proposed user selection method, the RMSE decreases to $0.0556$, resulting in a significant improvement in sensing accuracy. 

\begin{figure*}[t]
\centering
\subfigure[Image of a sunny day]{
\includegraphics[width=0.21\textwidth]{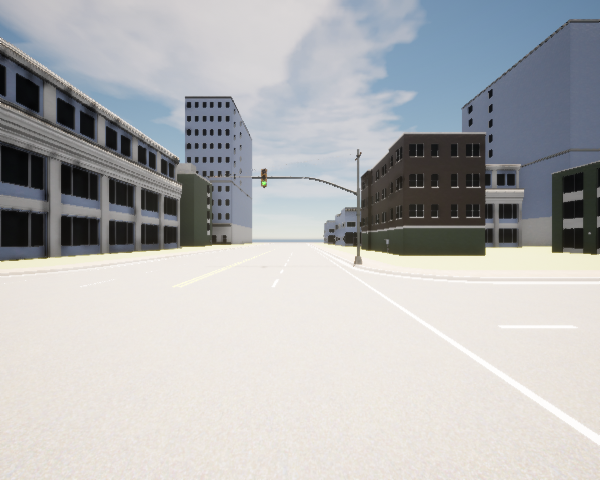}
}
\quad\quad
\subfigure[Image of a rainy day]{
\includegraphics[width=0.21\textwidth]{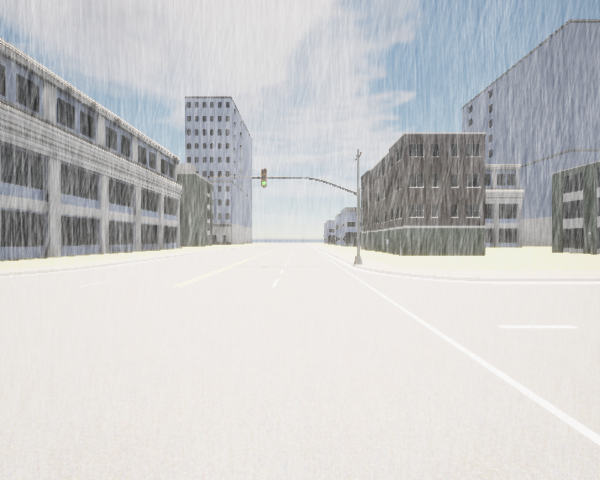}
}
\quad\quad
\subfigure[Image of a snowy day]{
\includegraphics[width=0.21\textwidth]{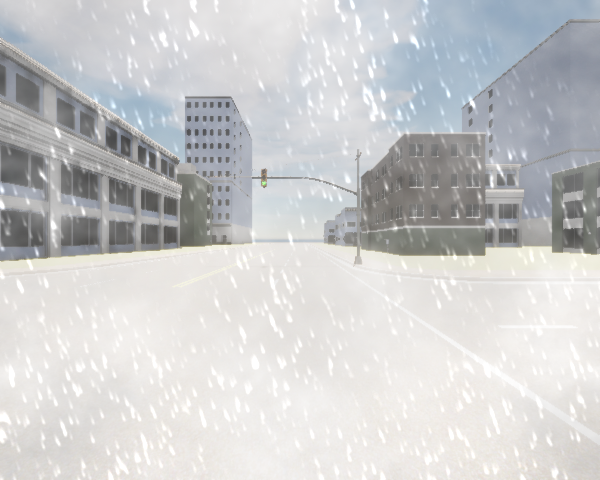}
}
\quad\quad\quad
\subfigure[Depth estimation by the image-only at a sunny day]{
\includegraphics[width=0.21\textwidth]{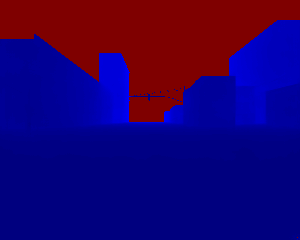}
}
\quad\quad
\subfigure[Depth estimation by the image-only at a rainy day]{
\includegraphics[width=0.21\textwidth]{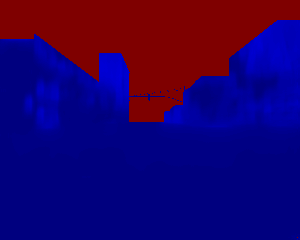}
}
\quad\quad
\subfigure[Depth estimation by the image-only at a snowy day]{
\includegraphics[width=0.21\textwidth]{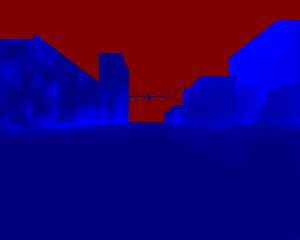}
}
\quad\quad\quad
\subfigure[Depth estimation by the proposed MMFN at a sunny day]{
\includegraphics[width=0.21\textwidth]{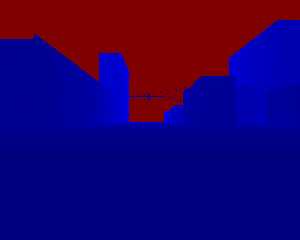}
}
\quad\quad
\subfigure[Depth estimation by the proposed MMFN at a rainy day]{
\includegraphics[width=0.21\textwidth]{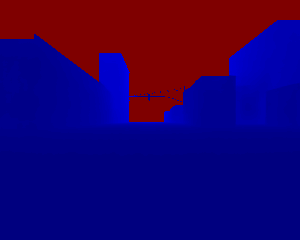}
}
\quad\quad
\subfigure[Depth estimation by the proposed MMFN at a snowy day]{
\includegraphics[width=0.21\textwidth]{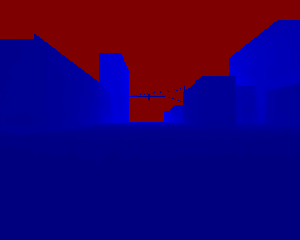}
}
\caption{The depth estimation under different weather conditions.}\label{pre_dpt}
\end{figure*}

In addition to the reconstruction of regular walls, as proposed in \cite{wang2023bayesian,yang2022hybrid}, we next attempt to reconstruct irregular walls.
However, in the environment with irregular walls, there may be blind spots when using one single BS for sensing.
As shown in Fig.~\ref{humian}, the wall in the bottom right corner cannot be sensed because there \textcolor[rgb]{0.00,0.00,0.00}{is no} first-order NLOS paths with reflection points located in the blind spots.
To eliminate the blind spots and obtain a complete environment reconstruction result, we utilize two BSs \textcolor[rgb]{0.00,0.00,0.00}{that} are placed on the left and right sides respectively for the sensing of the irregular walls.
Each BS first utilizes the proposed multi-user selection algorithm to select the superior users and generate the point set.
Since the point set obtained from each BS is relatively accurate, we then directly take the union of the two point sets from the two BSs and fit the curves as shown in Fig.~\ref{humian_two_bs}.
It can be seen that the multi-user selection based environment reconstruction can effectively reconstruct irregular walls as well.
Moreover, the calculated points from each BS all fall in the directions of the transmit beams from the BS, which is similar to the phenomenon in Fig.~\ref{proposed}.


Next, we evaluate the performance of the proposed multi-modal fusion based environment reconstruction under different weather conditions. Specifically, we consider fusing the image with the BS's point set by the proposed MMFN to accurately estimate the depth of each pixel within the image. We train the MMFN on the sunny dataset. The testing RMSE during the training process is shown in Fig.~\ref{trainloss}. The image-only method yields an RMSE of $5.281$, while the integration of communications sensing information and visual sensing information achieves an RMSE of $1.874$.
We further examine the robustness of the MMFN under rainy and snowy weather conditions.
The ground truth of the depth map is presented in Fig.~\ref{gt}. In the depth map, the varying shades of color represent the proximity of objects, where darker colors indicate closer distances and lighter colors represent farther distances. For the image-only method, the predicted depth map in a sunny environment is displayed in Fig. \ref{pre_dpt}(d), the predicted depth map in a rainy environment is displayed in Fig.~\ref{pre_dpt}(e), and the predicted depth map in a snowy environment is displayed in Fig.~\ref{pre_dpt}(f).
With the proposed MMFN, the predicted depth map in a sunny environment is displayed in Fig. \ref{pre_dpt}(g), the predicted depth map in a rainy environment is displayed in Fig. \ref{pre_dpt}(h), and the predicted depth map in a snowy environment is displayed in Fig. \ref{pre_dpt}(i).
It can be seen that the image-only method is comparable to the MMFN in sunny conditions.
However, in rainy and snowy conditions, the depth estimation performance of the image-only method is relatively poor, whereas the proposed MMFN still maintains a high accuracy.
Therefore, the performance of the image-only method and the proposed MMFN across different evaluation metrics is presented in TABLE \ref{table2}, where a downward arrow indicates smaller values are better while an upward arrow indicates larger values are better. In rainy conditions, the proposed MMFN achieves a depth estimation RMSE of $3.564$, which is approximately $2.9$ times better than the image-only method. Similarly, in snowy conditions, the proposed MMFN achieves an RMSE of $4.802$, which is approximately $3.8$ times better than the image only method. The proposed MMFN demonstrates strong robustness to different weather conditions.

\begin{figure}[t]
  \centering
  \includegraphics[width=0.5\textwidth]{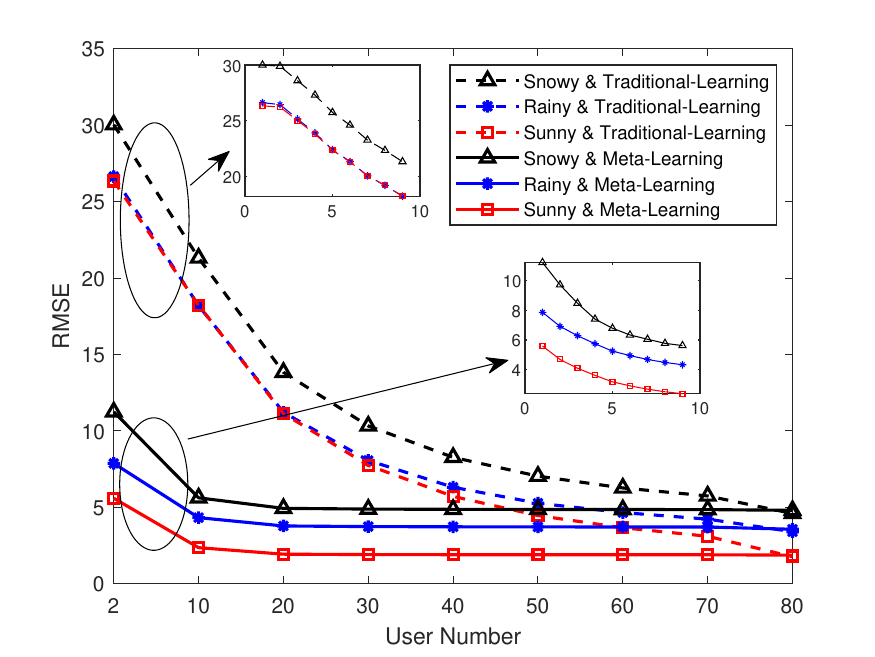}\\
  \caption{The RMSE when different users participate in the sensing task.}\label{user_num}
\end{figure}

To answer how many communications users are required for environment reconstruction, we next test the accuracy of depth estimation with different numbers of users. In Fig.~\ref{user_num}, the dashed line represents the MMFN trained using the traditional training strategy while the solid line represents the MMFN trained using the proposed meta-learning based strategy.
For the traditional training approach, the maximum user number is set to $N_{max}=80$.
During the training process, the input of the MMFN consists of the sensing data from $N_{max}$ communications users. However, during testing, when $u_t$ users' sensing data is used, the remaining $(N_{max}-ut)$ rows of the input are filled with $(0,0,1)$\footnote{The filled row is $(0,0,1)$, where $(0,0)$ represent the coordinate of the input image and the value $1$ represents the depth of the coordinate $(0,0)$ is the farthest. In other words, the filled $(0,0,1)$ represents the depth information of the top-left corner of the image, indicating the farthest distance (sky) information.}.
It can be seen from Fig. \ref{user_num} that the depth estimation error gradually decreases with the increasing of user number, which also demonstrates that the clustering characteristics of the communication system can bring significant benefits to ISAC. Moreover, the traditional training approach requires an increase in the number of users to $50$ before the performance reaches a satisfactory level. However, with the proposed meta-learning based training strategy, the performance converges to a satisfactory level after the number of users reaches $10$.
The proposed meta-learning based training strategy facilitates the activation of performance leaps, making it easier to trigger significant improvements as the number of users increases.

\section{Conclusions}\label{conclusion}

In this paper, we propose a multi-user based environment reconstruction scheme, where the BS collects the beam scanning information of the ubiquitous users to compute the environment point set. Moreover, we propose an evaluation criterion for sensing users and use this criterion to select users who can yield accurate reflection points. The point set is then distributed to users with sensing requests. Each user fuses the received point set with their own images by the proposed MMFN to achieve a more dense and comprehensive depth map.
Based on the user selection, we attained an impressive 16-fold enhancement in the precision of point set estimation.
Additionally, The proposed MMFN significantly enhances the accuracy of environment reconstruction under challenging weather conditions like \textcolor[rgb]{0.00,0.00,0.00}{raining and snowing}.
Moreover, we propose a meta-learning-based training approach that enables the network to be effective under any number of users, which greatly improves the deployability and feasibility of MMFN.

\bibliographystyle{IEEEtran}

\end{document}